\documentclass[final,10pt]{elsarticle}
\usepackage{etex}
\usepackage{amssymb,amsbsy,exscale,amsmath,amsfonts,amscd,amsthm}
\usepackage[letterpaper]{geometry}
\usepackage[usenames,dvipsnames]{color}
\usepackage{bm}
\usepackage{soul}
\usepackage[ampersand]{easylist}
\usepackage{graphicx}
\graphicspath{{./images/}}
\usepackage{parskip}
\usepackage{longtable}
\usepackage{subcaption}
\usepackage[usenames,dvipsnames]{color}
\usepackage{multirow}
\usepackage{algorithmicx}
\usepackage[ruled]{algorithm}
\usepackage{algpseudocode}
\biboptions{sort&compress}
\usepackage{pifont}
\usepackage{tikz}
\usepackage{epstopdf}
\usepackage[nottoc]{tocbibind}
\usetikzlibrary{mindmap,backgrounds}
\usepackage{stmaryrd}
\usepackage{mathrsfs}
\usepackage[format=hang,font=footnotesize]{caption}
\usepackage{epigraph}
\usepackage{array}
\usepackage{booktabs,makecell}
\usepackage{parskip}
\usepackage{mathtools}
\usepackage[mathscr]{euscript}
\usepackage{upgreek}
\usepackage[overload]{empheq}
\usepackage{xspace}
\usepackage{arydshln}
\usepackage{cmap}
\usepackage[T1]{fontenc}
\usepackage{url}
\newcommand{\lb}{\left[}
\newcommand{\rb}{\right]}
\newcommand{\lp}{\left(}
\newcommand{\rp}{\right)}

\DeclareCaptionLabelFormat{andtable}{#1~#2 \& \tablename~\thetable}
\newcommand\bigzero{\makebox(0,0){\text{\huge0}}}

\newcommand{\mbb}[1]{\mathbb{#1}}

\newcommand{\mcal}[1]{\mathcal{#1}}

\newcommand{\abs}[1]{\left\lvert{#1}\right\rvert}
\providecommand{\norm}[1]{\lVert#1\rVert}

\newcommand{\curl}{\textrm{curl}}
\newcommand{\dive}{\textrm{div}}

\newcommand{\el}{K}

\newcommand{\metdom}{\Omega^{\rm m}}
\newcommand{\didom}{\Omega^{\rm d}}
\newcommand{\mcT}[1]{\mathcal{T}^{\rm #1}}
\newcommand{\bx}{{\bf x}}

\newcommand{\bn}{{\bf n}}

\newcommand{\bt}{{\bf t}}

\newcommand{\bE}{{\bf E}}
\newcommand{\bH}{{\bf H}}
\newcommand{\bV}{{\bf V}}
\newcommand{\bW}{{\bf W}}
\newcommand{\bJ}{{\bf J}}

\newcommand{\bU}{{\bf U}}

\newcommand{\bv}{{\bf v}}

\newcommand{\bu}{{\bf u}}

\newcommand{\mm}{$\upmu$m\xspace}

\newcommand{\pow}{\uptau}

\makeatletter
\newcommand*\rel@kern[1]{\kern#1\dimexpr\macc@kerna}
\newcommand*\widebar[1]{%
  \begingroup
  \def\mathaccent##1##2{%
    \rel@kern{0.8}%
    \overline{\rel@kern{-0.8}\macc@nucleus\rel@kern{0.2}}%
    \rel@kern{-0.2}%
  }%
  \macc@depth\@ne
  \let\math@bgroup\@empty \let\math@egroup\macc@set@skewchar
  \mathsurround\z@ \frozen@everymath{\mathgroup\macc@group\relax}%
  \macc@set@skewchar\relax
  \let\mathaccentV\macc@nested@a
  \macc@nested@a\relax111{#1}%
  \endgroup
}
\makeatother

\newcommand{\COMSOL}{\textsc{Comsol }{Multiphysics}\xspace}

 \def\etal{{\it et al.}\xspace}

\def\ie{{\it i.e.}\xspace}

\usepackage[most]{tcolorbox}

\newcommand\revone[1]{\textcolor{black}{#1}}
\newcommand\revtwo[1]{\textcolor{black}{#1}}

\journal{Journal of Computational Physics}

\begin{document}

\begin{frontmatter}


\title{A nested hybridizable discontinuous Galerkin method for computing second-harmonic generation in three-dimensional metallic nanostructures}

\author[mit]{F.~Vidal-Codina\corref{cor1}}
\ead{fvidal@mit.edu}
\author[mit]{N.-C.~Nguyen}
\ead{cuongng@mit.edu}
\author[iit]{C.~Cirac\`{i}}
\ead{ciraci@iit.edu}
\author[umn]{S.-H.~Oh}
\ead{sang@umn.edu}
\author[mit]{J.~Peraire}
\ead{peraire@mit.edu}

\cortext[cor1]{Corresponding author}
\address[mit]{Department of Aeronautics and Astronautics, Massachusetts Institute of Technology, Cambridge, MA 02139, USA}
\address[iit]{Center for Biomolecular Nanotechnologies, Istituto Italiano di Tecnologia, Via Barsanti 14, 73010 Arnesano (LE), Italy}
\address[umn]{Department of Electrical and Computer Engineering, University of Minnesota, Minneapolis, MN 55455, USA}

 \begin{abstract}

{We develop a nested hybridizable discontinuous Galerkin (HDG) method to numerically solve the Maxwell's equations coupled with a  hydrodynamic model for the conduction-band electrons in  metals. The HDG method leverages static condensation to eliminate the degrees of freedom of the approximate solution defined in the elements, yielding a linear system in terms of the degrees of freedom of the approximate trace defined on the element boundaries. This article presents a computational method that relies on a degree-of-freedom reordering such that the HDG linear system accommodates an additional static condensation step to eliminate a large portion of the degrees of freedom of the approximate trace, thereby yielding a much smaller linear system. For the particular metallic structures considered in this article, the resulting linear system obtained by means of nested static condensations is a block tridiagonal system, which can be solved efficiently. We apply the nested HDG method to compute second harmonic generation on a triangular coaxial periodic nanogap structure. This nonlinear optics phenomenon features rapid field variations and extreme boundary-layer structures that span a wide range of length scales. Numerical results show that the ability to  identify structures which exhibit resonances at $\omega$ and $2\omega$ is essential to excite the second harmonic response.}

 \end{abstract}
\begin{keyword}
Hybridizable discontinuous Galerkin method \sep Maxwell's equations \sep hydrodynamic model for metals \sep nonlinear plasmonics \sep nonlocal electrodynamics \sep second-harmonic generation
  \end{keyword}
  
\end{frontmatter}

\section{Introduction}
Nonlinear plasmonics \cite{boyd2003nonlinear,kauranen2012nonlinear,panoiu2018nonlinear,smirnova2016multipolar,butet2015optical} studies the effects that arise when nonlinear media in a plasmonic structure cause the polarization to depend nonlinearly on the electric field. These effects are inherently weak, but can be amplified by the collective excitation of conduction-band electrons that occur on metallic nanostructures, commonly known as plasmon resonances. These excitations produce strong near-field enhancements of the incident wave by confining light in small volumes whose critical length scales are several orders of magnitude smaller than the wavelength of light. The combination of nonlinear effects and localized plasmon resonances offers new opportunities for the generation and manipulation of light at the nanoscale. The most common nonlinear effect is second-harmonic generation (SHG), whereby two photons at the incident frequency interact to generate a single photon at twice the incident frequency. Within classical electromagnetics, SHG occurs when the polarization has a  quadratic dependence on the electric field. In common nonlinear optical materials, this quadratic dependence is provided by the absence of an inversion symmetry in the crystalline lattice, which generates an asymmetric response with respect to the applied electric field orientation. 
In contrast, plasmonic metals (e.g. gold, silver, copper and aluminum) are centrosymmetric and do not possess an intrinsic second-order susceptibility. 
A correct description of metal nonlinearities has to account for the complex dynamics of free-electrons \cite{Scalora:2010kd}.
It is well-known in fact that free-electrons provide both bulk and surface mechanisms for SHG \cite{Jha:1965te,Sipe:1980vz,Wang:2009ce}.
Surface contributions arise from locally broken centrosymmetry while bulk contributions are ascribed to convective and Lorentz-force interactions in the electron fluid \cite{Scalora:2010kd,ciraci2012second}. 
Contrarily to nonlinear crystals, the nonlinear response of metals becomes highly dependent on the device geometry \cite{klein2006second,Canfield:2007kz}. 
In fact, unless centrosymmetry is broken at the larger scale of the device structure, the second-harmonic fields stays highly localized and destructively interferes in the far-field, becoming too weak to be observed.  Plenty of research has been devoted to designing and fabricating shapes that enhance SHG, see \cite{kauranen2012nonlinear,panoiu2018nonlinear,butet2015optical,krasavin2018free} and the references therein.

The ability to accurately model and simulate nonlinear plasmonic phenomena requires computational capabilities that challenge traditional simulation techniques. From a modeling perspective, plasmonic phenomena can be described by Maxwell's equations coupled with a hydrodynamic model to account for the nonlocal effect of the conduction-band electrons in the metallic materials, which become relevant for sub-10 nanometer features. The problems of interest involve the interaction of long-wavelength electromagnetic waves (\mm and mm) with nanometer-wide features for potential applications in sensing and spectroscopy. Moreover, the electromagnetic fields are confined in deep-subwavelength cavities, and \AA ngstrom-thin accumulation charge layers develop at the metal-dielectric interface. As a consequence, the discretizations required to attain accurate simulations need to be highly adapted (to properly capture the extremely localized fields) and anisotropic (to be computationally tractable). 

The finite-difference time-domain (FDTD) algorithm \cite{taflove2005computational,kunz1993finite} is a well known computational method for wave propagation. The most popular FDTD method utilizes Yee's scheme \cite{yee1966numerical} to discretize space and time with staggered cartesian grids and second-order schemes. The main shortcoming of FDTD is modeling geometries with complex features, since the stair-casing at the interfaces not aligned with the Cartesian grid severely impacts the accuracy. In addition, the mismatch in length scales that is characteristic of nonlinear plasmonics is a severe hindrance, since resolving for the smallest phenomena while keeping a uniform discretization may require  grids having prohibitively  large number of degrees of freedom.  The FDTD method has been used for nonlocal \cite{mcmahon2010calculating,fang2016full} and second-harmonic generation \cite{zeng2009classical,celebrano2015mode,liu2010generalization,aouani2012multiresonant} simulations. 

Finite element (FE) methods \cite{jin2015finite} have also been widely used in electromagnetics, due to their ability to handle heterogeneous media and intricate geometries with unstructured discretizations, as well as $h/p$ adaptation for increased accuracy. The family of face/edge elements introduced by N\'{e}d\'{e}lec \cite{nedelec1980mixed} have been extensively used to simulate electromagnetic wave propagation, and have been shown to avoid the problem of spurious modes \cite{bossavit1990solving} by appropriately choosing the approximation spaces. A commonly used implementation of edge elements for Maxwell's equations is the one provided by the RF Module of \COMSOL \cite{comsol}. Many groups have used this platform to implement their own version of the hydrodynamic model \cite{toscano2012modified,ciraci2012probing,toscano2015resonance}, SHG on nanoparticles and nanoantennas \cite{bachelier2008multipolar,zhang2011three,berthelot2012silencing,carletti2015enhanced,ginzburg2015nonperturbative} as well as SHG on periodic arrays of nanostructures \cite{kolkowski2016non,ciraci2012second,chandrasekar2015second,razdolski2016resonant}.%

Discontinuous Galerkin (DG) methods \cite{cockburn1998local,hesthaven2002nodal} have also been widely used to simulate nonlocal effects in electromagnetics. In DG methods, the domain is discretized onto a collection of disjoint elements, and each field component is independently approximated within each element using standard finite element spaces. Solutions are therefore discontinuous across elements, and flux continuity is enforced at the interfaces. The DG method in time domain has been developed for many nanophotonics applications on metallic and lossless media \cite{busch2011discontinuous,niegemann2009simulation,lu2004discontinuous,ji2007high,lanteri2013convergence,schmitt2016dgtd}, as well as to simulate SHG on nanoparticles \cite{kullock2011shg,hille2016second,moeferdt2018plasmonic}, resonators and antennas \cite{grynko2017simulation,von2013isolated,linden2012collective,alberti2016role}. The main caveat of DG methods for 3-D applications in the frequency domain or in the time domain with implicit time integration is the high computational burden, stemming from the duplication of degrees of freedom at the interfaces. This shortcoming is circumvented by the hybridizable discontinuous Galerkin (HDG) method, first developed in \cite{cockburn2009unified,cockburn2008superconvergent,cockburn2009superconvergent} and later extended to acoustics and elastodynamics \cite{nguyen2011acoustic,saa2012binary} as well as time-harmonic Maxwell's equations \cite{nguyen2011maxwell,li2013hybridizable} and the hydrodynamic model for metals \cite{vidal2018hybridizable,li2017hybridizable}. In addition, unlike other DG methods HDG exhibits optimal convergence rates for both the solution and the flux. As a result, the solution may be locally post-processed to gain an additional order of convergence, a phenomenon known as superconvergence. In the recent years, the HDG method has been successfully applied for 2-D and 3-D metallic nanostructures to simulate plasmonic phenomena \cite{park2015nanogap,yoo2016high,vidal2018computing,vidal2018hybridizable,li2017hybridizable,yoo2019modeling,vidal2020terahertz}.


This article presents a nested hybridizable discontinuous Galerkin (nHDG) method for the Maxwell's equations coupled with a hydrodynamic model for the conduction-band electrons in metals. By means of a static condensation to eliminate the degrees of freedom of the approximate solution defined within the elements, the HDG method yields a linear system in terms of the degrees of freedom of the approximate trace defined on the element boundaries. The nested HDG is a computational method that builds on top of classical HDG, whereby an additional static condensation is performed in order to eliminate a large portion of the degrees of freedom of the HDG linear system. Consequently, this nested strategy gives rise to a much smaller linear system, encompassing the degrees of freedom of a reduced number of element boundaries. Furthermore, if these element boundaries are judiciously selected, the nested HDG yields a linear system that is block-tridiagonal, and can thus be solved efficiently. \revtwo{The article also presents the formulation and implementation of the HDG method to compute SHG under the assumption of non-depleted pump approximation, that is, the fundamental wave is not affected by the generated harmonic \cite{boyd2003nonlinear}}. We apply the nested HDG method to compute the SHG on a triangular coaxial periodic nanogap structure, which is a computationally intensive task since nonlinear optics phenomena feature rapid field variations and extreme boundary-layer structures that span a wide range of length scales. In addition, we propose strategies to partition the mesh in order to achieve an efficient nested static condensation.

This article is organized as follows. In Section \ref{sec:problem}, we introduce the equations and notation used throughout the article. In Section \ref{sec:hdg}, we review the formulation and implementation of the HDG method to solve the hydrodynamic model for metals in frequency domain, describe the modifications needed to simulate SHG and present the nested hybridization strategy and algorithm. In Section \ref{sec:res}, we present numerical results to assess the performance of the proposed method and present some concluding remarks in Section \ref{sec:conc}.

\section{Second harmonic generation in metallic nanostructures}\label{sec:problem}
We now derive the equations that will be used throughout this article. The overline $\overline{\phantom{x}}$ denotes dimensional  variables and constants, whereas the quantities without overlines are their non-dimensional counterparts. The only dimensional quantities without overlines are the reference quantities that we use to non-dimensionalize the problem: $L_{\rm c}$ is a reference length scale in meters, $\alpha$ is a reference magnetic field in Ampere/meter, $\varepsilon_0$ is the free-space permittivity  in Farad/meter, $\mu_0$ is the free-space permeability in Henry/meter, $c_0 = 1/\sqrt{\varepsilon_0\mu_0}$ is the free-space speed of light in meters/second and $Z_0 = \sqrt{\mu_0/\varepsilon_0}$ is the free-space impedance in Volt/Ampere.

\subsection{Maxwell's equations in time domain}

The electric $\widebar{\mcal{E}}(\widebar{\bx},\widebar{t})$ and magnetic $\widebar{\mcal{H}}(\widebar{\bx},\widebar{t})$ fields, along with the electric displacement $\widebar{\mcal{D}}$ and magnetic flux density $\widebar{\mcal{B}}$, satisfy Maxwell's equations in a metallic domain $\metdom$
\begin{equation}\label{eq:maxwell_s}
 \begin{aligned}
\widebar{\nabla} \times \widebar{\mcal{E}} + \partial_{\widebar{t}}\widebar{\mcal{B}}&= 0 \quad \mbox{(Amp\`{e}re's law)}, \\
\widebar{\nabla} \times \widebar{\mcal{H}} -\partial_{\widebar{t}} \widebar{\mcal{D}}&=  \widebar{\mcal{J}}_{\textnormal{ext}} \quad \mbox{(Faraday's law)},\\
\widebar{\nabla}\cdot \widebar{\mcal{D}}&= \widebar{\rho}_{\textnormal{ext}},\quad \mbox{(Gauss's law)},\\
\widebar{\nabla} \cdot \widebar{\mcal{B}}& = 0, \quad \mbox{(magnetic Gauss's law)},
\end{aligned}
\end{equation}
where $\widebar{\mcal{J}}_{\textnormal{ext}} $ represents the external electric current and $\widebar{\rho}_{\textnormal{ext}} $ the external volume charge density. For simplicity of exposition, we assume there are no external current $\widebar{\mcal{J}}_{\textnormal{ext}} = 0$ and external charge density $\widebar{\rho}_{\textnormal{ext}} = 0$. In addition, we have the following constitutive relations
\begin{equation}\label{eq:constitutive_s}
 \begin{aligned}
\widebar{\mcal{B}} &= \widebar{\mu} \widebar{\mcal{H}}\,,\\
\widebar{\mcal{D}} &= \varepsilon_0\widebar{\mcal{E}} + \widebar{\mcal{P}}+ \widebar{\mcal{P}}_\infty = \widebar{\varepsilon}_\infty\widebar{\mcal{E}} + \widebar{\mcal{P}}\,, \\
\widebar{\nabla} \cdot \widebar{\mcal{P}} &= -\widebar{\rho}\,\\
 \partial_{\widebar{t}}\widebar{\mcal{P}}&= \widebar{\mcal{J}}  \,.
\end{aligned}
\end{equation}
The polarization density $\widebar{\mcal{P}}$ represents the density of permanent or induced electric dipole moments due to free electrons. Conversely, the background polarization $\widebar{\mcal{P}}_\infty = (\widebar{\varepsilon}_\infty-\varepsilon_0)\widebar{\mcal{E}}$ represents the polarization of the bound electrons in the valence band. The last two relations relate the polarization density $\widebar{\mcal{P}}$ to the internal current $\widebar{\mcal{J}}$ and internal charge density $\widebar{\rho}$. 


\subsection{A nonlinear hydrodynamic model}

The above set of equations is closed once we specify the polarization density of the material in response to applied electromagnetic fields. In the simplest case, $\widebar{\mcal{P}}$ depends locally on the electric field through a linear relationship. To account for nonlocal effects which become important at nanoametric scales, a hydrodynamic model (HM) for the free electron gas was proposed in \cite{eguiluz1975hydro}. This model, despite neglecting quantum phenomena such as quantum tunneling and quantum oscillations, introduces a hydrodynamic pressure term that accounts for the nonlocal coupling of the conduction-band electrons. Below we provide a brief description of the model and refer to \cite{boardman1982electromagnetic,pitarke2006theory,ciraci2013hydrodynamic} for additional details.

The electron density $\widebar{n}(\widebar{\bx},\widebar{t})$ and the hydrodynamic velocity $\widebar{\bf v}(\widebar{\bx},\widebar{t})$ are related by the continuity equation as $\partial_{\widebar{t}} \widebar{n}= -\widebar{\nabla}\cdot (\widebar{n}\widebar{\bv})$. In addition, the equation of motion for the electron fluid under a macroscopic electromagnetic field is described as
\begin{equation}\label{eq:nonlocal0}
 \widebar{m}_e(\partial_{\widebar{t}} +\widebar{\bv}\cdot \widebar{\nabla} + \widebar{\gamma})\widebar{\bv} = \widebar{e}(\widebar{\mcal{E}} + \widebar{\bv}\times\widebar{\mcal{B}}) - \frac{\widebar{\nabla} \widebar{p}}{\widebar{n}}\;,
\end{equation}
where $m_e$ is the effective electron mass, $e$ is the electron charge so that $\widebar{\mcal{J}} = \widebar{e}\widebar{n}\widebar{\bv}$ and $\widebar{\gamma}$ is a damping constant related to the collision rate of the electrons. The electron pressure $\widebar{p}(\widebar{\bx},\widebar{t})$ is given by \cite{crouseilles2008quantum} 
\begin{equation}\label{pressure}
p(\widebar{\bx},\widebar{t}) = \widebar{p}_0 \left(\frac{\widebar{n}(\widebar{\bx},\widebar{t})}{\widebar{n}_0}\right)^{5/3}
\end{equation}
where $\widebar{p}_0 =\dfrac{18}{25} \widebar{n}_0 \widebar{E}_{\rm F}$, $\widebar{E}_{\rm F}$ is the Fermi energy and $\widebar{n}_0$ is the equilibrium charge density.

After combining equations  \eqref{eq:nonlocal0},  \eqref{pressure} and the continuity equation, we follow \cite{ciraci2012second} and retaining only first- and second-order terms, we obtain a nonlinear nonlocal model for the polarization in response to an applied electromagnetic field
\begin{equation}\label{eq:hm_time}
\begin{split}
-\widebar{\beta}^2 \widebar{\nabla}(\widebar{\nabla}\cdot \widebar{\mcal{P}}) + \partial_{\widebar{t}\widebar{t}}\widebar{\mcal{P}} +\widebar{\gamma}\partial_{\widebar{t}}\widebar{\mcal{P}} -\widebar{\omega}_{\rm p}^2\varepsilon_0 \widebar{\mcal{E}}= & \; - \dfrac{  \widebar{\omega}_{\rm p}^2 \varepsilon_0}{\widebar{n}_0 \widebar{e}}\widebar{\mcal{E}}\lp\widebar{\nabla}\cdot \widebar{\mcal{P}} \rp + \dfrac{  \widebar{\omega}_{\rm p}^2}{c_0^2 \widebar{n}_0 \widebar{e}}\partial_{\widebar{t}}\widebar{\mcal{P}}\times \widebar{\mcal{H}}\\ &- \dfrac{1}{\widebar{n}_0 \widebar{e}} \widebar{\nabla}\cdot(\partial_{\widebar{t}}\widebar{\mcal{P}} \otimes \partial_{\widebar{t}}\widebar{\mcal{P}}) - \dfrac{ \widebar{\beta}^2}{3 \widebar{n}_0 \widebar{e}} \widebar{\nabla} (\widebar{ \nabla}\cdot \widebar{\mcal{P}})^2 \;,
\end{split}
\end{equation}
 where $\widebar{\omega}_{\rm p} = \sqrt{\widebar{n}_0 \widebar{e}^2/\widebar{m}_e \varepsilon_0}$ is the metal's plasma frequency. For incoming fields above this frequency the metal behaves like a lossy dielectric since electron mobility is not sufficient to react and cancel the incoming wave. The nonlocal parameter $\widebar{\beta}$ is given by  $\widebar{\beta} = \sqrt{6 \widebar{E}_{\rm F}/ 5 \widebar{m}_e}$. Given the definition of Fermi kinetic energy $\widebar{E}_{\rm F} = \widebar{m}_e \widebar{v}_{\rm F}^2/2$, the nonlocal parameter reduces to $\widebar{\beta} = \sqrt{3/5} \, \widebar{v}_{\rm F}$ \cite{lindhard1954properties}, where $\widebar{v}_{\rm F}$ is the Fermi velocity. The nonlinear terms in the right hand side of equation (\ref{eq:hm_time}) are  known as Coulomb force, magnetic Lorentz force,  nonlinear convective force  and nonlinear pressure force. We note that the simplest form of the hydrodynamic model, based on the Thomas-Fermi approximation accounting only for the linearized kinetic energy, corresponds to neglecting the nonlinear terms in the right hand side of \eqref{eq:hm_time} to recover a linear expression.
 The implementation of this nonlocal linear model with a hybridizable discontinuous Galerkin method was presented in \cite{vidal2018hybridizable}.
 
 Equation \eqref{eq:hm_time} needs to be solved simultaneously with Maxwell's equations \eqref{eq:maxwell_s}. Before proceeding, it is convenient to non-dimensionalize the problem variables using the following scalings
\begin{equation}\label{eq:scalings}
 \begin{aligned}
\widebar{\bx} &=L_{\rm c} \bx ,\quad \widebar{t} = L_{\rm c}t /c_0 ,\quad \widebar{\mcal{E}} = \alpha Z_0 \mcal{E}, \quad  \widebar{\mcal{H}} = \alpha \mcal{H},\\
\widebar{\mcal{D}} &= \varepsilon_0 \alpha Z_0 \mcal{D},\quad \widebar{\mcal{B}} = \mu_0 \alpha \mcal{B},\quad \widebar{\mcal{J}} =  \alpha \mcal{J}/L_{\rm c}, \quad  \widebar{\mcal{P}} =  \alpha \mcal{P}/c_0\;.
\end{aligned}
\end{equation}
For a non-magnetic medium $(\widebar{\mu} = \mu_0)$, applying the scalings above to Maxwell's equations \eqref{eq:maxwell_s}, the constitutive relations \eqref{eq:constitutive_s} and the hydrodynamic pressure equation \eqref{eq:hm_time}, we obtain 
\begin{equation}\label{eq:nondim_eqs}
 \begin{aligned}
\nabla \times \mcal{E} + \partial_t\mcal{H}&= 0 ,\\
\nabla \times \mcal{H} -\partial_t (\varepsilon_\infty\mcal{E})&= \mcal{J} , \\
{\beta}^2 \nabla\rho + \partial_{{t}}{\mcal{J}} + {\gamma}{\mcal{J}} -\omega_{\rm p}^2{\mcal{E}}&=  \dfrac{\omega_{\rm p}^2  }{ n_0 e} {\mcal{E}} \rho + \dfrac{ \omega_{\rm p}^2 }{ n_0 e } {\mcal{J}}\times {\mcal{H}} - \dfrac{1 }{n_0 e } {\nabla}\cdot({\mcal{J}} \otimes {\mcal{J}}) -  \dfrac{ {\beta}^2 }{3 n_0 e} \nabla \rho^2 ,
\end{aligned}
\end{equation}
with the non-dimensional optical constants $\varepsilon_\infty = \widebar{\varepsilon}_\infty/\varepsilon_0$, $\omega_{\rm p} =\widebar{\omega}_{\rm p}L_{\rm c}/c_0$, $\gamma=\widebar{\gamma}L_{\rm c}/c_0$, $\beta = \widebar{\beta}/c_0$ and electron constants $e = \widebar{e}\,c_0/(\alpha L^2_{\rm c})$ and $n_0 = \widebar{n}_0 L^3_{\rm c}$.

We now assume a time-harmonic response of the non-dimensional electromagnetic fields, where $^*$ denotes the complex conjugate
\begin{equation}\label{eq:harmonic}
\begin{aligned}
\mcal{E}(\bx,t) &= \sum_{n\in \mathbb{N}}  \Re\lbrace\bE_n(\bx)\exp(-in\omega t) \rbrace = \dfrac{1}{2}\sum_{n\in \mathbb{N}}  \bE_n(\bx)\exp(-in\omega t) + {\bE}^*_n(\bx)\exp(in\omega t)\;, \\
\mcal{H}(\bx,t) &= \sum_{n\in \mathbb{N}}  \Re\lbrace\bH_n(\bx)\exp(-in\omega t) \rbrace = \dfrac{1}{2}\sum_{n\in \mathbb{N}}  \bH_n(\bx)\exp(-in\omega t) + {\bH}^*_n(\bx)\exp(in\omega t)\;, \\
\mcal{J}(\bx,t) &= \sum_{n\in \mathbb{N}}  \Re\lbrace\bJ_n(\bx)\exp(-in\omega t) \rbrace = \dfrac{1}{2}\sum_{n\in \mathbb{N}}  \bJ_n(\bx)\exp(-in\omega t) + {\bJ}^*_n(\bx)\exp(in\omega t)\;,
\\
\mcal{\rho}(\bx,t)  &= \sum_{n\in \mathbb{N}}  \Re\lbrace\uprho_n(\bx)\exp(-in\omega t) \rbrace = \dfrac{1}{2}\sum_{n\in \mathbb{N}}  \uprho_n(\bx)\exp(-in\omega t) + {\uprho}^*_n(\bx)\exp(in\omega t)\;.
\end{aligned}
\end{equation}
\revtwo{To derive a system for the amplitudes of the first harmonic, we introduce the expansions \eqref{eq:harmonic} into \eqref{eq:nondim_eqs} considering only $n \le 2$, multiply by $\exp \lp-i\omega t \rp$ and invoke orthogonality of the Fourier modes to obtain following set of nonlinear equations}
\begin{equation}\label{eq:maxwellhydro}
 \begin{aligned}
\nabla \times \bE_1 -i\omega\bH_1 &= {\bf 0} \,,\\
\nabla \times \bH_1 + i\omega\varepsilon_\infty \bE_1- \bJ_1&= {\bf 0}\, , \\
\beta^2 \nabla\uprho_1 + (\gamma - i\omega )\bJ_1 - \omega_{\rm p}^2\bE_1&= \revtwo{{\bm f}_1}\,,\\
i\omega\uprho_1 - \nabla\cdot \bJ_1 &= 0\,.
\end{aligned}
\end{equation}
\revtwo{The nonlinear source term $\bm f_1$ is given by
\begin{equation}\label{eq:source1}
\begin{split}
 2\bm f_1 &=  \dfrac{\omega_{\rm p}^2  }{n_0e} \lp\bE^*_1\uprho_2 + \bE_2\uprho^*_1\rp + \dfrac{ \omega_{\rm p}^2 }{ n_0e} \lp \bJ^*_1\times\bH_2 + \bJ_2\times\bH^*_1\rp \\
 &- \dfrac{1 }{n_0e} \nabla\cdot \lp \bJ^*_1 \otimes\bJ_2 + \bJ_2 \otimes\bJ^*_1 \rp - \dfrac{ {\beta}^2 }{3 n_0e} \nabla \lp\uprho^*_1\uprho_2 + \uprho_2\uprho^*_1\rp \;. 
\end{split}
\end{equation}
Since we expect the power of the first harmonic to be several orders of magnitude stronger than that of the second harmonic ($\norm{\bE_2}\ll \norm{\bE_1}$), we hereafter assume that $\bm f_1 = \bm 0$. This approximation is known as the non-depleted pump approximation \cite{boyd2003nonlinear}, which emphasizes that the second harmonic does not deplete the fundamental wave. This assumption simplifies the calculation of the second-harmonic and is justified in instances where the efficiency of the second harmonic generation is sufficiently weak.
For later use, we refer to the system of equations (\ref{eq:maxwellhydro}) with the simplification $\bm f_1 = \bm 0$,  as ${\mcal{L}^{\rm m}}(\bE_1, \bH_1, \bJ_1, \uprho_1; \omega) = {\bf 0}$.}

The Drude model \cite{drude1900elektronentheorie} can be recovered from \eqref{eq:maxwellhydro} by setting the nonlocal parameter $\beta$ to zero, in which case  Ohm's law is recovered $\bJ_1 = i\omega_{\rm p}^2\bE_1/(\omega+i\gamma)$ and the complex Drude permittivity is expressed as $\varepsilon_{\rm m}(\omega) = \varepsilon_\infty - \omega_{\rm p}^2/(\omega(\omega + i\gamma))$. The Drude model is more computationally efficient, since (\ref{eq:maxwellhydro}) simplifies to Maxwell's equations with a complex-valued permittivity, at the expense of neglecting the nonlocal electron interactions that become relevant for sub-10 nm features.

\subsection{Metallic nanostructures}
The above formulation is extended to consider the more general case of a metallic nanostructure, comprising both a metal $\metdom$ described by the HM and a dielectric $\didom$ with permittivity $\varepsilon_{\rm d}$ described solely by Maxwell's equations. The solution within the metallic structure is governed by (\ref{eq:maxwellhydro}) and (\ref{eq:maxwellhydro_BC}),  whereas the response in the dielectric $\didom$ is given by regular time-harmonic Maxwell's equations, namely
\begin{equation}\label{eq:maxwelld}
 \begin{aligned}
\nabla \times \bE_1 -i\omega\bH_1 &= {\bf 0} \,,\\
\nabla \times \bH_1 + i\omega\varepsilon_{\rm d} \bE_1&= {\bf 0} \,.
\end{aligned}
\end{equation}
The boundary conditions for a metallic nanostructure can be expressed as
\begin{equation}\label{eq:maxwellhydro_BC}
 \begin{aligned}
&\bn\times\bE_1\times\bn  = {\bf 0}, \quad  \mbox{on }\partial\Omega_{\rm{E}}\,,\\
&\bn\times\bH_1 =  {\bf 0},\quad  \mbox{on }\partial\Omega_{\rm{H}}\,,\\
&\bn\cdot\bJ_1  = 0, \quad \mbox{on }\partial\metdom_{\rm{J}} \cup \Gamma_{\rm md}\,,\\
&\uprho_1  = 0, \quad \mbox{on }\partial\metdom_{\uprho}\,,\\
&\bH_1\times \bn -  \sqrt{\varepsilon_{\rm d}}\,\bn\times\bE_1\times\bn = \bH_{\rm inc}\times \bn -  \sqrt{\varepsilon_{\rm d}}\,\bn\times\bE_{\rm inc}\times\bn := {\bm f}_{\rm inc},\quad \mbox{on }\partial\didom_{\rm{rad}}\,.
\end{aligned}
\end{equation}
The first and second boundary condition prescribe perfect electric conductor (PEC) and perfect magnetic conductor (PMC) behavior, which allows us to impose symmetries in periodic structures. The third and fourth conditions also prescribe symmetry conditions in periodic structures for the electric current and the electron charge, and are only applicable on the metallic subdomain. In addition, the third boundary condition is also applied at the metal-dielectric interface $\Gamma_{\rm md} = \metdom \cap \didom$ to preclude the electrons from leaving the metal (since the normal component of the electric current vanishes), also known as no electron spill-out condition \cite{boardman1976surface}. Quantum effects such as electron tunneling are therefore not modeled by the HM. 

The last equation is the first-order Silver-M\"{u}ller radiation condition \cite{sommerfeld1949partial,mur1981absorbing}, preventing outgoing waves from reflecting at the computational boundary and coming back into the domain. A common alternative to the radiation condition are the perfectly matched layers (PMLs) \cite{berenger1994perfectly,johnson2008notes}. Even though PMLs can be more effective at absorbing waves, they are more computationally intensive and require parameter tuning. We have developed both alternatives and found no significant differences for the metallic nanostructures considered within the frequency regimes of interest, hence we resorted to Silver-M\"{u}ller conditions. Illumination is prescribed as a ${\bf p}$-polarized plane wave propagating in the ${\bf d}$-direction, that is ${\bf E}_{\rm inc} = {\bf p}\exp(i\omega\sqrt{\varepsilon_{\rm d}}\,{\bf d}\cdot \bx)$ and ${\bf H}_{\rm inc} =  -i/\omega \nabla\times \bE_{\rm inc}$. For an example of how the boundary conditions are assigned see Fig. \ref{fig:triangle} (c), where a periodic triangular coaxial nanostructure is shown.

We refer to the above system of equations \eqref{eq:maxwelld} to be solved on the dielectric as  ${\mcal{L}^{\rm d}}(\bE_1, \bH_1; \omega) = {\bf 0}$, and the boundary equations \eqref{eq:maxwellhydro_BC} to be prescribed as ${{b}}(\bE_1, \bH_1, \bJ_1, \uprho_1; \omega) = {\bm f}_{\rm inc}$, respectively. In order to numerically solve the above systems with the HDG method we shall also impose continuity of the tangential component of the magnetic field along the entire domain and continuity of the normal component of the electric current along the metal. These two additional conditions are explained and derived within the HDG discretization, see Section \ref{sec:hdg}.

\subsection{Second-harmonic generation}
Once the solution for the fundamental harmonic $(\bE_1,\bH_1,\bJ_1,\uprho_1)$ has been determined by simultaneously solving \eqref{eq:maxwellhydro}, \eqref{eq:maxwelld} and \eqref{eq:maxwellhydro_BC}, we turn our attention to the second harmonic. \revtwo{Similarly as before, we introduce expansions \eqref{eq:harmonic} with $n \le2$ into \eqref{eq:nondim_eqs}, multiply by $\exp(-2i\omega t)$, invoke orthogonality of the Fourier basis and obtain the following set of equations for the amplitudes of the second harmonic in the metallic domain.
}.

\begin{equation}\label{eq:2nd}
\begin{aligned}
\nabla \times \bE_2 -2i\omega\bH_2 &= {\bf 0} ,\\
\nabla \times \bH_2 + 2i\omega\varepsilon_\infty \bE_2 - \bJ_2&= {\bf 0} , \\
\beta^2 \nabla\uprho_2 + (\gamma - 2i\omega)\bJ_2 - \omega_{\rm p}^2\bE_2&= {\bm f_2}\\
2i\omega\uprho_2 - \nabla\cdot \bJ_2 &= 0,
\end{aligned}
\end{equation}
where
\begin{equation} \label{eq:nlsource}
\begin{aligned}
 2\bm f_2 &=  \dfrac{\omega_{\rm p}^2  }{n_0e} \bE_1\uprho_1 + \dfrac{ \omega_{\rm p}^2 }{ n_0e} \bJ_1\times\bH_1 -  \dfrac{1 }{n_0 e} \nabla\cdot \lp \bJ_1 \otimes\bJ_1 \rp - \dfrac{ {\beta}^2 }{3  n_0e } \nabla \uprho_1^2
\end{aligned}
\end{equation} 
The nonlinear source term in \eqref{eq:nlsource} depends only on the fundamental fields and represents the motion of the electron fluid under an electromagnetic field. 

\revtwo{The assumption of non-depleted pump approximation simplifies second-harmonic calculations, since instead of solving a coupled system for the two harmonics, the SHG may be obtained by sequentially solving Maxwell's equations first for the fundamental wave and then for the second harmonic with a nonlinear source term involving only the fundamental fields \cite{ciraci2012second}.} Using the operator ${\mcal{L}^{\rm m}}$ introduced earlier, we can express equation (\ref{eq:2nd}) as ${\mcal{L}^{\rm m}}(\bE_2, \bH_2, \bJ_2, \rho_2 ;2\omega) = {\bm f_2}$. The boundary conditions for the SHG are then simply  $b(\bE_2, \bH_2, \bJ_2, \rho_2 ;2\omega) = {\bf 0}$, since no incident light is shone and the only response is due to the nonlinear current.

Summarizing, the non-depleted SHG can be simulated as a two-step process, namely
\begin{enumerate}
\item Solve 
 \begin{equation}\label{eq:fhg}
 \begin{aligned}
&\mcal{L}^{\rm m}(\bE_1, \bH_1, \bJ_1, \rho_1; \omega) = {\bf 0}\;, & \mbox{ in } \Omega^{\rm m}\;,\\
& \mcal{L}^{\rm d}(\bE_1, \bH_1; \omega) = {\bf 0}\;, & \mbox{ in } \Omega^{\rm d}\;,\\
&\mbox{with } \;{b}(\bE_1, \bH_1, \bJ_1, \uprho_1;\omega) = {\bm f}_{\rm inc}\;, &
 \end{aligned}
\end{equation}
to compute $(\bE_1,\bH_1,\bJ_1,\uprho_1)$.
\item Solve 
 \begin{equation}\label{eq:shg}
 \begin{aligned}
&\mcal{L}^{\rm m}(\bE_2, \bH_2, \bJ_2, \uprho_2; 2\omega) = {\bm f}_2(\bE_1,\bH_1,\bJ_1,\uprho_1)\;, & \mbox{ in } \Omega^{\rm m}\;,\\
& \mcal{L}^{\rm d}(\bE_2, \bH_2; 2\omega) = {\bf 0}\;, & \mbox{ in } \Omega^{\rm d}\;,\\
&\mbox{with } \;{b}(\bE_2, \bH_2, \bJ_2, \uprho_2;2\omega) = {\bf 0}\;, &
 \end{aligned}
\end{equation}
to compute $(\bE_2,\bH_2,\bJ_2,\uprho_2)$.
\end{enumerate}
Here, the plane-wave illumination is used only to compute the fundamental fields, ensuring the second harmonic fields are solely generated by the nonlinear source \eqref{eq:nlsource}. \revtwo{We point out that if the non-depleted approximation were not justified, the nonlinear source term \eqref{eq:source1} would need to be retained in \eqref{eq:fhg} and in this case, a simple fixed-point iterative algorithm involving \eqref{eq:fhg}  and \eqref{eq:shg} could be devised.}


\section{Nested HDG method for the hydrodynamic model}\label{sec:hdg}

\subsection{Approximation spaces}

We first review the notation, operators and approximation spaces needed for the HDG method following \cite{nguyen2011maxwell}. We denote by $\mcT{} = \mcT{m}\cup\mcT{d}$ a triangulation of disjoint regular elements $\el$ that partition a nanostructure consisting of a metallic and dielectric subdomains $\mcal{D}=\metdom\cup\didom\in\mbb{R}^3$. The set of element boundaries is then defined as $\partial \mcT{}:=\lbrace \partial \el:\,\el\in\mcT{} \rbrace$. For an arbitrary element $\el\in\mcT{}$, $F = \partial \el\cap \partial \mcal{D}$ is a boundary face if it has a non-zero 2-D Lebesgue measure. Any pair of elements $\el^+$ and $\el^-$ share an interior face $F = \partial \el^+ \cap \partial \el^-$ if its 2-D Lebesgue measure is non-zero. We finally denote by $\mcal{E}_h^o$ and $\mcal{E}_h^\partial$ the set of interior and boundary faces, respectively, and their union by $\mcal{E}_h = \mcal{E}_h^o\cup \mcal{E}_h^\partial$.

Let $\bn^+$ and $\bn^-$ be the outward-pointing unit normal vectors on the neighboring elements $\el^+,\,\el^-$, respectively. We further use $\bu^\pm$ to denote the trace of $\bu$ on $F$ from the interior of $\el^\pm$. The jump $\llbracket\cdot\rrbracket$ for an interior face $F\in\mcal{E}_h^o$ is defined as
\begin{equation*}
 \llbracket \bu \odot \bn \rrbracket = \bu^+\odot\bn^+ + \bu^-\odot\bn^-,
\end{equation*}
and for a boundary face $F\in\mcal{E}_h^\partial$ with outward normal $\bn$ as
\begin{equation*}
 \llbracket \bu \odot \bn \rrbracket = \bu\odot\bn.
\end{equation*}
Here, the binary operation $\odot$ represents either $\cdot$ or $\times$. The tangential $\bu^t$ and normal $\bu^n$ components of $\bu$, such that $\bu =\bu^t + \bu^n$, are given by
\begin{equation*}
 \bu^t= \bn \times\bu\times\bn\;,\qquad \bu^n = \bn(\bu\cdot\bn)\;.
\end{equation*}

Let $\bm L^2(\mcal{D})\equiv [L^2(\mcal{D})]^3$ denote the Lebesgue space of three dimensional square integrable vector functions and $H^1(\mcal{D})$ the Hilbert space  $H^1(\mcal{D}) = \lbrace v\in L^2(\mcal{D}):\,\int_\mcal{D}\abs{\nabla v}^2 <\infty\rbrace$. We introduce the curl-conforming space
\begin{equation*}
 \bm H^{\curl}(\mcal{D}) = \lbrace \bu \in \bm L^2(\mcal{D}):\nabla\times \bu \in \bm L^2(\mcal{D}) \rbrace
\end{equation*}
with associated norm $\norm{\bu}^2_{ \bm H^{\curl}(\mcal{D})} = \int_{\mcal{D}} \abs{\bu}^2 + \abs{\nabla\times\bu}^2$, as well as the div-conforming space
\begin{equation*}
 \bm H^{\dive}(\mcal{D}) = \lbrace \bu \in \bm L^2(\mcal{D}):\nabla\cdot \bu \in L^2(\mcal{D}) \rbrace
\end{equation*}
with associated norm $\norm{\bu}^2_{ \bm H^{\dive}(\mcal{D})} = \int_\mcal{D} \abs{\bu}^2 + \abs{\nabla\cdot\bu}^2$. 

Let $\mcal{P}^p(\mcal{D})$ denote the space of complex-valued polynomials of degree at most $p$ on $\mcal{D}$. We introduce the following approximation spaces 
\begin{align*}
W_h &= \{w\in L^2(\mcal{D}) : w|_{\el} \in \mcal{P}^{p}(\el), \;\forall \el \in \mathcal{T}_h\},\\
\bm W_h &= \{\bm w \in \bm L^2(\mcal{D}) : \bm w|_{\el} \in \lb \mcal{P}^{p}(\el) \rb^3,\; \forall \el\in \mathcal{T}_h\},\\
M_h &= \{\mu\in  L^2(\mcal{E}_h)\,: \mu|_{F} \in \mcal{P}^{p}(F),\;\forall F\in \mcal{E}_h \},\\
\bm M_h &= \{\bm\mu\in \bm L^2(\mcal{E}_h)\,: \bm\mu|_{F} \in \mcal{P}^{p}(F)\bt_1 \oplus \mcal{P}^{p}(F)\bt_2,\;\forall F\in \mcal{E}_h \},
\end{align*}
where $\bt_1,\, \bt_2$ are linearly independent vectors tangent to the face. We note that by construction, $\bm\mu \in \bm M_h$ satisfies $\bm\mu = \bn\times \bm\mu\times\bn = \mu_1 \bt_1 + \mu_2 \bt_2$. The tangent vectors on a face $F$ can be defined in terms of its normal $\bn = (n_1,n_2,n_3$) as ${\bf t}_1 =(-n_2/n_1,1,0)$ and ${\bf t}_2 =(-n_3/n_1,0,1)$. This definition assumes that $|n_1| \ge \max(|n_2|,|n_3|)$ but analogous expressions can be obtained when $|n_2| \ge \max(|n_1|,|n_3|)$ or $|n_3| \ge \max(|n_1|,|n_2|)$ to avoid singularities. Boundary conditions are included by setting $\bm M_h ({\bf u}_\partial\vert_{\partial \mcal{D}}) = \lbrace \bm \mu \in \bm M_h:\, \bn\times\bm\mu\times\bn = \Pi {\bf u}_\partial\; \mbox{on } \partial \mcal{D} \rbrace$ and $ M_h ({ u}_\partial\vert_{\partial \mcal{D}}) = \lbrace  \mu \in  M_h:\, \mu = \Pi { u}_\partial\; \mbox{on } \partial \mcal{D} \rbrace$, where $\Pi {\bf u}_\partial$ (respectively, $\Pi u_\partial$) is the projection of the prescribed value of ${\bf u}$, ${\bf u}_\partial$, onto $\bm M_h$ (respectively, $u_\partial$ onto $M_h$).

Finally, we define the various Hermitian products for the above finite element spaces.  The volume inner products are defined as
\begin{equation*}
 (\eta,\zeta)_{\mcT{}} := \sum_{\el\in\mcT{}}(\eta,\zeta)_{\el},\qquad  (\bm\eta,\bm\zeta)_{\mcT{}} := \sum_{i = 1}^3(\eta_i,\zeta_i)_{\mcT{}},
\end{equation*}
and the surface inner products by
\begin{equation*}
 \langle\eta,\zeta\rangle_{\partial\mcT{}} := \sum_{\el\in\mcT{}} \langle\eta,\zeta \rangle_{\partial \el},\qquad   \langle\bm\eta,\bm\zeta \rangle_{\partial\mcT{}} := \sum_{i = 1}^3 \langle \eta_i,\zeta_i \rangle_{\partial\mcT{}}.
\end{equation*}
For two arbitrary scalar functions $\eta$ and $\zeta$, its scalar product $(\eta,\zeta)_\mcal{D}$ is the integral of $\eta\zeta^*$ on $\mcal{D}$. 

\subsection{First hybridization}

In this section, we describe the HDG discretization for a metallic nanostructure, introduced in \cite{vidal2018hybridizable}, which is a necessary first step to develop the nested HDG method. For completeness, we will consider both plane-wave illumination -- active only when computing first harmonic -- as well as the nonlinear source term -- active only when computing second harmonic. 

The HDG discretization of $\mcal{L}^{\rm m},\, \mcal{L}^{\rm d}$ and $b$ needs to be completed with two additional continuity condition, that is enforcing zero jump in the tangential component of $\bH_h$ and in the normal component of $\bJ_h$. For all test functions $(\bm \kappa,\bm\eta,\bm \xi,\zeta,\bm\mu,\theta)\in \bm W_h \times \bm W_h \times \bm W_h \times W_h \times \bm M_h \times M_h$, we seek approximate fields $(\bH_h,\bE_h,\bJ_h,\uprho_h,\widehat{\bE}_h,\widehat{\uprho}_h) \in \bm W_h \times \bm W_h \times \bm W_h \times W_h \times \bm M_h ({\bf 0 }\vert_{\partial \Omega_{\rm{E}}})  \times M_h({ 0 }\vert_{\partial \metdom_{\uprho}})$ such that
\begin{equation}\label{eq:hdg_hydro1}
 \begin{aligned}
-i\omega(\bH_h,\bm \kappa)_{\mcT{m}} + (\bE_h,\nabla\times \bm \kappa)_{\mcT{m}} + \langle \widehat{\bE}_h,\bm \kappa\times\bn \rangle_{\partial{\mcT{m}} \backslash \partial\metdom_{\rm{E}}} & = {\bf 0}, \\
-\beta^2 (\uprho_h,\nabla\cdot\bm\eta)_{\mcT{m}} + \beta^2\langle \widehat{\uprho}_h,\bm\eta\cdot\bn \rangle_{\partial{\mcT{m}}\backslash \partial\metdom_{\uprho}} + (\gamma - i\omega )(\bJ_h,\bm\eta)_{\mcT{m}} - \omega_{\rm p}^2(\bE_h,\bm\eta)_{\mcT{m}}&= ({\bm f},\bm\eta)_{\mcT{m}},\\
(\bH_h,\nabla \times \bm \xi)_{\mcT{m}}+ \langle \widehat{\bH}_h,\bm \xi\times\bn \rangle_{\partial{\mcT{m}}} +i \omega(\varepsilon_\infty\bE_h,\bm \xi)_{\mcT{m}} -(\bJ_h,\bm \xi)_{\mcT{m}}  & = {\bf 0} ,\\
i\omega(\uprho_h,\zeta)_{\mcT{m}} - \langle \widehat{\bJ}_h\cdot \bn,\zeta \rangle_{\partial{\mcT{m}}} + (\bJ_h,\nabla \zeta)_{\mcT{m}} &= 0, \\
-\langle \bn\times \widehat{\bH}_h,\bm\mu \rangle_{\partial{\mcT{m}}\backslash \partial \metdom_{\rm E}} &={\bf 0}, \\
\langle \widehat{\bJ}_h \cdot \bn ,\theta \rangle_{\partial{\mcT{m}}\backslash \partial \metdom_{\uprho}}&= 0, 
\end{aligned} 
\end{equation}
are satisfied in the metal. The fields $\widehat{\bE}_h,\,\widehat{\bH}_h,\,\widehat{\bJ}_h,\,\widehat{\uprho}_h$ are single valued on the faces and approximate the tangential component of $\bE,\,\bH,\,\bJ$ and  the trace of $\uprho$, respectively. The first four equations are the weak formulation of  ${\mcal{L}^{\rm m}}(\bH,\bE,\bJ,\uprho; \omega) = {{\bm f}_{\rm inc}}$. The boundary equations are strongly prescribed on the approximation spaces for the electric field and electron charge and weakly prescribed on the last two equations for the magnetic field and the electric current. In addition, the fifth equation enforces zero jump in the tangential component of $\bH_h$, that is $\llbracket \bn\times \widehat{\bH}_h \rrbracket = {\bf 0}$ along all elemental interfaces $\mcal{E}_h$, and lastly the sixth equation enforces zero jump on the normal component of $\bJ_h$ along all metal-metal interfaces $\mcal{E}^{\rm m}_h$, that is $\llbracket \bn\cdot \widehat{\bJ}_h \rrbracket = {0}$.

Similarly, for the dielectric domain $\didom$ the following weak formulation is satisfied
\begin{equation}\label{eq:hdg_max1}
 \begin{aligned}
-i\omega(\bH_h,\bm \kappa)_{\mcT{d}} + (\bE_h,\nabla\times \bm \kappa)_{\mcal{T}_h} + \langle \widehat{\bE}_h,\bm \kappa\times\bn \rangle_{\partial{\mcT{d}} \backslash \partial\didom_{\rm{E}}} & = {\bf 0}, \\
(\bH_h,\nabla \times \bm \xi)_{\mcT{d}}+ \langle \widehat{\bH}_h,\bm \xi\times\bn \rangle_{\partial{\mcT{d}}} +i \omega(\varepsilon_{\rm d}\bE_h,\bm \xi)_{\mcT{d}}  & ={\bf 0} ,\\
-\langle \bn\times \widehat{\bH}_h,\bm\mu \rangle_{\partial{\mcT{d}} \backslash \partial \didom_{\rm E}}  - \sqrt{\varepsilon_{\rm d}}\langle \widehat{\bE}_h,\bm\mu \rangle_{ \partial \didom_{\rm rad}}&= \langle {\bm f}_{\rm inc},\bm\mu \rangle_{ \partial \didom_{\rm rad}} \\
\end{aligned} 
\end{equation}

We close the system by introducing expressions for the hybrid fluxes of the magnetic field and electric current field as 
\begin{equation}
\begin{aligned}\label{eq:traces}
 \widehat{\bH}_h &= {\bH}_h + \tau_t(\bE_h - \widehat{\bE}_h)\times\bn,\\
  \widehat{\bJ}_h\cdot \bn &= {\bJ}_h\cdot \bn - \tau_ni\omega(\uprho_h-\widehat{\uprho}_h).
\end{aligned} 
\end{equation}
The parameters $\tau_t,\,\tau_n$ are the stabilization parameters, defined globally to ensure the accuracy and stability of the HDG discretization. We propose the choice $\tau_t =\sqrt{\varepsilon}$ (for each material's permittivity) and $\tau_n = \omega_{\rm p}/\beta$ which leads to numerically stable solutions even in the presence of tightly localized fields in the metal-dielectric interface. 

Substituting  \eqref{eq:traces} in \eqref{eq:hdg_hydro1} and integrating by parts, we write the final HDG discretization for the metallic domain
\begin{equation}\label{eq:hdg_hydro}
 \begin{aligned}
-i\omega(\bH_h,\bm \kappa)_{\mcT{m}} + (\bE_h,\nabla\times \bm \kappa)_{\mcT{m}} + \langle \widehat{\bE}_h,\bm \kappa\times\bn \rangle_{\partial{\mcT{m}} \backslash \partial\metdom_{\rm{E}}} & = {\bf 0}, \\
-\beta^2 (\uprho_h,\nabla\cdot\bm\eta)_{\mcT{m}} + \beta^2\langle \widehat{\uprho}_h,\bm\eta\cdot\bn \rangle_{\partial{\mcT{m}}\backslash \partial\metdom_{\uprho}} + (\gamma - i\omega )(\bJ_h,\bm\eta)_{\mcT{m}} - \omega_{\rm p}^2(\bE_h,\bm\eta)_{\mcT{m}}&= ({\bm f}_2,\bm\eta)_{\mcT{m}},\\
(\nabla \times \bH_h,\bm \xi)_{\mcT{m}} +\langle  \tau_t[\bE_h-\widehat{\bE}_h],\bn\times\bm \xi\times\bn \rangle_{\partial\mcT{m}} + i\omega(\varepsilon_\infty\bE_h,\bm \xi)_{\mcT{m}} - (\bJ_h,\bm \xi)_{\mcT{m}}  &={\bf 0} , \\
- (\nabla\cdot {\bJ}_h,\zeta )_{\mcT{m}} + i\omega(\uprho_h,\zeta)_{\mcT{m}}  + i\omega \tau_n\langle  \uprho_h,\zeta \rangle_{\partial\mcT{m}} - i\omega \tau_n\langle\widehat{\uprho}_h,\zeta \rangle_{\partial\mcT{m}} &= 0,\\
-\langle \bn\times {\bH}_h ,\bm\mu \rangle_{\partial\mcT{m} \backslash \partial \metdom_{\rm E}}-\langle \tau_t \bE_h,\bm\mu \rangle_{\partial\mcT{m} \backslash \partial \metdom_{\rm E}} +\langle{\tau}_t \widehat{\bE}_h,\bm\mu \rangle_{\partial\mcT{m} \backslash \partial \metdom_{\rm E}} &= {\bf 0}, \\
\langle {\bJ}_h \cdot \bn ,\theta \rangle_{\partial\mcT{m} \backslash\partial \metdom_{\uprho}}  - i\omega\tau_n \langle \uprho_h ,\theta \rangle_{\partial\mcT{m} \backslash\partial \metdom_{\uprho}} + i\omega\tau_n \langle\widehat{\uprho}_h ,\theta \rangle_{\partial\mcT{m} \backslash\partial \metdom_{\uprho}}&= 0.
\end{aligned} 
\end{equation}
The nonlinear source also requires integration by parts, hence
\begin{equation*}
\begin{split}
({\bm f_2},\bm\eta)_{\mcT{m}} &= \dfrac{\omega_{\rm p}^2  }{2n_0e } \lp\bE_1\uprho_1,\bm\eta\rp_{\mcT{m}} + \dfrac{ \omega_{\rm p}^2 }{2n_0e} \lp \bJ_1\times\bH_1\rp_{\mcT{m}} \\
&-   \dfrac{1 }{2n_0e} \lb \langle ( \widehat{\bJ}_1 \otimes\widehat{\bJ}_1 ) \cdot \bn,\bm\eta\rangle_{\partial\mcT{m}} - \lp {\bJ}_1 \otimes{\bJ}_1,\nabla\bm\eta\rp_{\mcT{m}} \rb \\
&- \dfrac{  {\beta}^2 }{6n_0 e}\lb \langle \widehat{\uprho}_1^2,\bm\eta\cdot\bn\rangle_{\partial\mcT{m}} - (\uprho_1^2,\nabla\cdot\bm\eta)_{\mcT{m}}  \rb\;.
\end{split}
\end{equation*}
The final HDG discretization for the dielectric is obtained after substituting
\eqref{eq:traces} in \eqref{eq:hdg_max1} and integrating by parts
\begin{equation}\label{eq:hdg_max}
 \begin{aligned}
-i\omega(\bH_h,\bm \kappa)_{\mcT{d}} + (\bE_h,\nabla\times \bm \kappa)_{\mcT{d}} + \langle \widehat{\bE}_h,\bm \kappa\times\bn \rangle_{\partial{\mcT{d}} \backslash \partial\didom_{\rm{E}}} & = {\bf 0}, \\
(\nabla \times \bH_h,\bm \xi)_{\mcT{d}} +\langle  \tau_t[\bE_h-\widehat{\bE}_h],\bn\times\bm \xi\times\bn \rangle_{\partial\mcT{d}} + i\omega(\varepsilon_{\rm d}\bE_h,\bm \xi)_{\mcT{d}}   &={\bf 0} , \\
\begin{split}
 -\langle \bn\times {\bH}_h ,\bm\mu \rangle_{\partial\mcT{d} \backslash \partial \didom_{\rm E}}-\langle \tau_t \bE_h,\bm\mu
\rangle_{\partial\mcT{d} \backslash \partial \didom_{\rm E}}+&\\ +\langle{\tau}_t \widehat{\bE}_h,\bm\mu \rangle_{\partial\mcT{d} \backslash \partial \didom_{\rm E}}
- \sqrt{\varepsilon_{\rm d}}\langle \widehat{\bE}_h,\bm\mu \rangle_{ \partial \didom_{\rm rad}}&= \langle{\bm f}_{\rm inc},\bm\mu\rangle_{ \partial \didom_{\rm rad}}.   
\end{split}
\end{aligned} 
\end{equation}

 
The weak formulations \eqref{eq:hdg_hydro}-\eqref{eq:hdg_max} are then discretized using the corresponding basis functions on all the elements and faces of $\mcT{}$, thus giving rise to the linear system 
\renewcommand{\arraystretch}{1.5}
\begin{equation}\label{eq:hdg_metal}
\left[\begin{array}{cccc:cc}
-i\omega\mbb{A} & 0& \mbb{B} & 0 &\mbb{C} & 0\\
0& (\gamma-i\omega)\mbb{A} & -\omega_{\rm p}^2\mbb{A} & -\beta^2\mbb{P} &0 & \beta^2\mbb{O}\\
\mbb{B}^T & -\mbb{A} & \mbb{D} +i\omega\mbb{A}_\varepsilon & 0 & -\mbb{E} &0 \\
0 & -\mbb{P}^T& 0& i\omega\mbb{H} & 0 &-i\omega\mbb{N}\\
\hdashline
\mbb{C}^T& 0& -\mbb{E}^T & 0 &\mbb{M} &0\\
0 & \mbb{O}^T& 0& -i\omega\mbb{N}^T & 0 &i\omega\mbb{T}\\
\end{array}\right] \left[\begin{array}{c} \underline{\bH}_{\rm m} \\ \underline{\bJ}_{\rm m} \\ \underline{\bE}_{\rm m} \\ \underline{\uprho}_{\rm m}\\ \hdashline \underline{\widehat{\bE}}_{\rm m}\\ \underline{\widehat{\uprho}}_{\rm m}\end{array} \right] =  \left[\begin{array}{c} { 0}\\ \underline{\bm f_2}\\ { 0} \\ { 0} \\ \hdashline{ 0}\\ { 0}\end{array} \right]\;, 
\end{equation}
for the metal domain, and 
\renewcommand{\arraystretch}{1.5}
\begin{equation}\label{eq:hdg_dielectric}
\left[\begin{array}{cc:c}
-i\omega\mbb{A} & \mbb{B} & \mbb{C} \\
\mbb{B}^T & \mbb{D} +i\omega\mbb{A}_\varepsilon & -\mbb{E} \\
\hdashline
\mbb{C}^T& -\mbb{E}^T & \mbb{M}\\
\end{array}\right] \left[\begin{array}{c} \underline{\bH}_{\rm d}\\ \underline{\bE}_{\rm d} \\ \hdashline\underline{\widehat{\bE}}_{\rm d}\end{array} \right] =  \left[\begin{array}{c} { 0}\\ { 0} \\ \hdashline\underline{\bm f_{\rm inc}} \end{array} \right],
\end{equation}
for the dielectric domain. Here $(\underline{\bH}_{\rm m}, \underline{\bE}_{\rm m}, \underline{\widehat{\bE}}_{\rm m})$ and $(\underline{\bH}_{\rm d}, \underline{\bE}_{\rm d}, \underline{\widehat{\bE}}_{\rm d})$ represent the vectors of degrees of freedom of $(\bH_h, \bE_h, \widehat{\bE}_h)$ in metal and dielectric, respectively. Similarly, $(\underline{\bJ}_{\rm m}, \underline{\uprho}_{\rm m}, \underline{\widehat{\uprho}}_{\rm m})$ represent the vectors of degrees of freedom of $(\bJ_h, {\uprho}_h, \widehat{\uprho}_h)$ in metal. It is important to point out that the two systems (\ref{eq:hdg_metal}) and (\ref{eq:hdg_dielectric}) have to be solved simultaneously because $\underline{\widehat{\bE}}_{\rm m}$ and $\underline{\widehat{\bE}}_{\rm d}$ share degrees of freedom for the faces located on the metal-dielectric interface.

We note that due to the discontinuous nature of the approximation spaces, the local variables $\bE_h,\,\bH_h,\,\bJ_h,\,\uprho_h$ (defined in the interior of each element) are only coupled globally through the global variables $\widehat{\bE},\,\widehat{\uprho}$ (defined on the element faces). This means that we can eliminate these local variables at the element level, which in matrix form corresponds to eliminating the upper-left submatrices, indicated with the dashed lines in (\ref{eq:hdg_metal}) and (\ref{eq:hdg_dielectric}), thus only a reduced matrix for the global variables needs to be assembled. This numerical strategy, also known as hybridization or static condensation, is essential to achieve an efficient implementation of the HDG method \cite{cockburn2009unified,cockburn2008superconvergent,cockburn2009superconvergent,nguyen2009linearCD,nguyen2011acoustic,nguyen2011maxwell}. Specifically, the hybridization procedure yields the following global linear system
\begin{equation}\label{eq:globalsystem}
\bm\Upsilon \widehat{\bU} = {\bm b}, 
\end{equation}
where the vector $\widehat{\bU}$ consists of the degrees of freedom of $(\widehat{\bE}_h, \widehat{\uprho}_h)$. In practice, both the matrix $\bm\Upsilon$ and the vector $\bm b$ are formed by a standard finite element assembly procedure by computing the elemental quantities and assembling them in an element-by-element fashion. The detailed implementation can be found in \cite{vidal2018hybridizable,vidal2017simulation}.

The elimination of local degrees of freedom through hybridization renders a linear system where the global degrees of freedom are defined on the faces only, thus drastically reducing the size of the linear system that must be solved.  After solving the linear system (\ref{eq:globalsystem}) for the global unknowns, the local unknowns can be efficiently recovered at the element level \cite{vidal2018hybridizable,vidal2017simulation}, an operation that is trivially parallelizable.

\subsection{Nested hybridization}
Thus far, we have recreated the formulation and implementation of the HDG method for a metal-dielectric domain introduced in \cite{vidal2018hybridizable}. In this section, we describe a nested hybridization method to efficiently solve the global linear system that stems from the first hybridization \eqref{eq:globalsystem} by exploiting the geometry of the problems of interest. Indeed, for large 3-D structures the direct solution of \eqref{eq:globalsystem} may be challenging. Iterative methods, on the other hand, have found limited success for stiff indefinite problems of the type considered here.

The idea behind the nested hybridization is to partition the global degrees of freedom $\widehat{\bU} = \lbrace {\bW},\,{\bV} \rbrace$, such that the ${\bW}$ degrees of freedom can be statically condensed to yield the following linear system 
\begin{equation}
\label{Vlinearsystem}
\bm\Psi  {\bV} =  {\bm d}    
\end{equation}
for ${\bV}$. Specifically, reordering the global system \eqref{eq:globalsystem} using the  $\widehat{\bU} = \lbrace {\bW},\,{\bV} \rbrace$ allows us to write of $\bm\Psi,\,{\bm d}$ as a function of  $\bm\Upsilon,\,{\bm b}$, namely
 \renewcommand{\arraystretch}{1.75}
\begin{equation}
\left[\begin{array}{c:c}
\bm\Upsilon_{{\rm W}{\rm W}} &  \bm\Upsilon_{{\rm W}{{\rm V}}}  \\
\hdashline
\bm\Upsilon_{{{\rm V}}{\rm W}} & \bm\Upsilon_{{{\rm V}}{{\rm V}}}
\end{array}\right] \left[\begin{array}{c} {\bW}\\\hdashline {\bV}\end{array} \right] = \left[\begin{array}{c} {\bm b}_{{\rm W}} \\\hdashline {\bm b}_{ {\rm V}} \end{array} \right] . 
\end{equation}
Assuming that $\bm\Upsilon_{{{\rm W}}{\rm W}}$ is invertible, the above system may be recast as 
\begin{eqnarray}
{\bW}  & = \bm\Upsilon_{{\rm W}{\rm W}}^{-1}\lb  {\bm b}_{{\rm W}} - \bm\Upsilon_{{\rm W}{\rm V}} {\bV} \rb  \label{eq:hybridization1} \\
 \lb  \bm\Upsilon_{{\rm V}{\rm V}} -  \bm\Upsilon_{{\rm V}{\rm W}}  \bm\Upsilon_{{\rm W}{\rm W}}^{-1} \bm\Upsilon_{{\rm W}{\rm V}}\rb  {\bV} & =  {\bm b}_{{\rm V}} -  \bm\Upsilon_{{\rm V}{\rm W}} \bm\Upsilon_{{\rm W}{\rm W}}^{-1}{\bm b}_{{\rm W}} \;. \label{eq:hybridization2}
\end{eqnarray}
Thus we have  $\bm\Psi = \bm\Upsilon_{{\rm V}{\rm V}} -  \bm\Upsilon_{{\rm V}{\rm W}}  \bm\Upsilon_{{\rm W}{\rm W}}^{-1} \bm\Upsilon_{{\rm W}{\rm V}}$ and $\bm d = {\bm b}_{{\rm V}} -  \bm\Upsilon_{{\rm V}{\rm W}} \bm\Upsilon_{{\rm W}{\rm W}}^{-1}{\bm b}_{{\rm W}}$.


For this hybridization to be computationally efficient, we need to ensure the inverse of $\bm\Upsilon_{{{\rm W}}{\rm W}}$ may be efficiently evaluated by means of a judicious choice of the global degrees of freedom. 
\begin{figure}[h!]
 \centering
 \includegraphics[scale = .9]{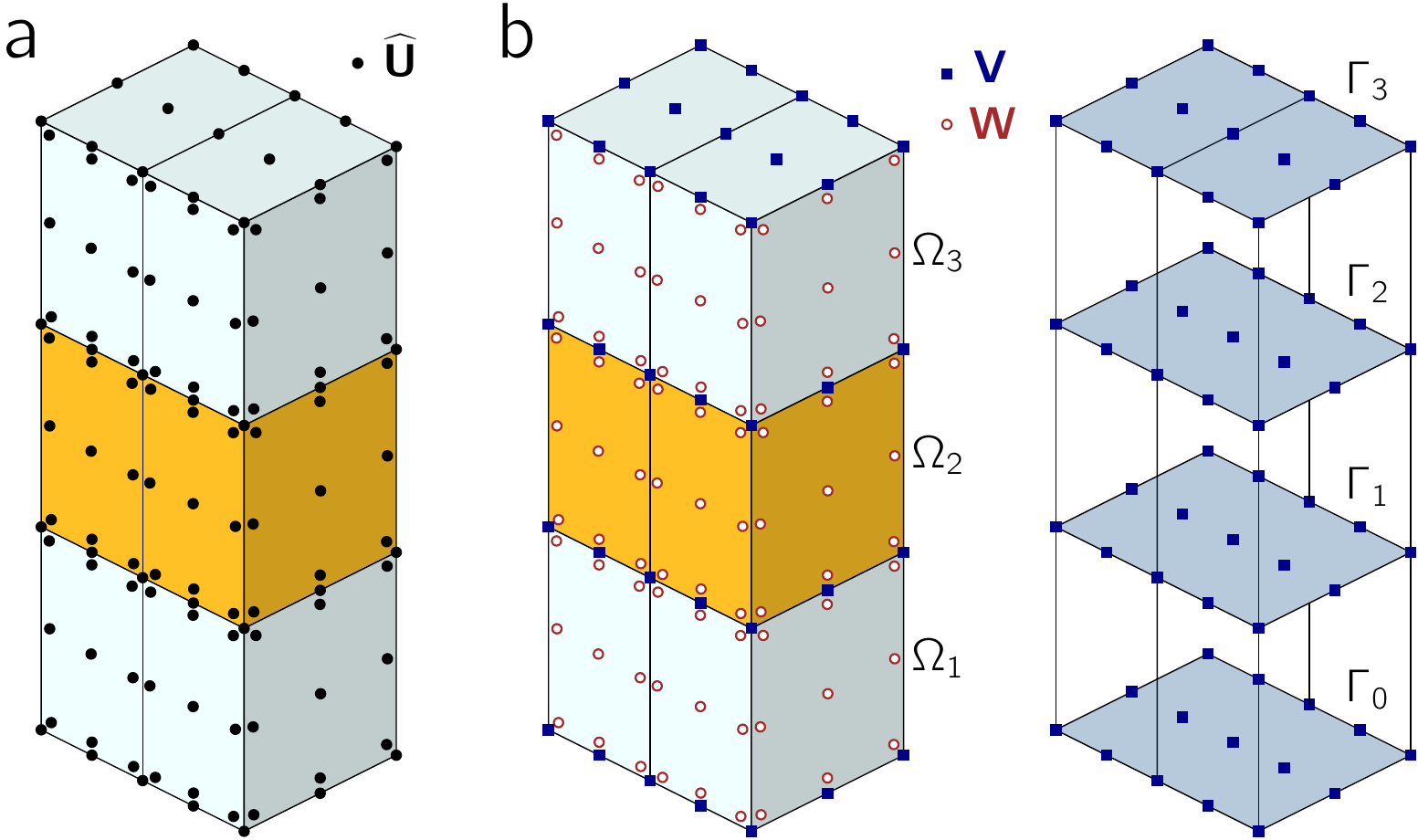}
 \caption{Sample metallic film structure with degrees of freedom for a $p=2$ discretization. (a) Global degrees of freedom $\widehat{\bU}$ after first hybridization. (b) Degree of freedom splitting $\widehat{\bU} = \lbrace {\bW},\,{\bV} \rbrace$ after nested hybridization, with ${\bV}$ only on horizontal planes and ${\bW}$ within the blocks.}\label{fig:2hybrid}
\end{figure}
To that end, we target specifically metallic nanostructures that can be discretized by extruding a 2-D discretization. There are many examples of such structures in the literature \cite{zhu2016quantum,ciraci2012probing,chen2013atomic,yoo2016high,yoo2018high,ciraci2012second,vidal2018hybridizable,park2015nanogap}.  Under this assumption, the degrees of freedom in the planes perpendicular to the extrusion direction are assigned to ${\bV}$, and the remaining degrees of freedom are assigned to ${\bW}$. This partitioning gives rise to $\rm N$ blocks $\lbrace \Omega_n\rbrace_{n=1}^{\rm N}$, whose unknowns are $\lbrace{\bW}_n\rbrace_{n=1}^{\rm N}$; and $\rm N+1$ interfaces $\lbrace \Gamma_n\rbrace_{n=0}^{\rm N}$, whose unknowns are $\lbrace{\bV}_n\rbrace_{n=0}^{\rm N}$, as illustrated in Fig. \ref{fig:2hybrid}. Consequently, the matrix $\bm\Upsilon_{{{\rm W}}{\rm W}}$ can be inverted efficiently since it is a block-diagonal matrix, namely 
$$\bm\Upsilon_{{{\rm W}}{\rm W}} = \mathrm{diag} \left( \bm\Upsilon_{{{\rm W}_1}{\rm W}_1}, \bm\Upsilon_{{{\rm W}_2}{\rm W}_2}, \ldots, \bm\Upsilon_{{{\rm W}_{\rm N}}{\rm W}_{\rm N}} \right) . $$
Furthermore, this partitioning gives rise to a linear system \eqref{Vlinearsystem}  that is a block-tridiagonal, namely
\begin{equation}\label{eq:tridiagonal}
 \lb\begin{array}{cccccc}
\bm\alpha_0 & \bm\beta_0 &  &  & & \\
\bm\gamma_0 & \bm\alpha_1 & \bm\beta_1 &  &\bigzero &\\
& \bm\gamma_1 & \ddots & \ddots & & \\
&  & \ddots & \ddots & \ddots& \\
&  \bigzero& & \ddots &  \bm\alpha_{\rm N -1}& \bm\beta_{\rm N -1}\\
 & & && \bm\gamma_{\rm N -1} & \bm\alpha_{\rm N} \end{array} \rb\;  \left[\begin{array}{c}{\bf V}_0 \\ {\bf V}_1 \\ \vdots\\ \vdots\\{\bf V}_{\rm N -1} \\ {\bf V}_{\rm N} \end{array} \right]  =  \left[\begin{array}{c}{\bm d}_0 \\ {\bm d}_1 \\ \vdots\\ \vdots\\{\bm d}_{\rm N -1} \\ {\bm d}_{\rm N} \end{array} \right] 
\end{equation}
The blocks may be computed explicitly following \eqref{eq:hybridization1} and \eqref{eq:hybridization2}  to obtain
\begin{equation}
\begin{aligned}\label{eq:blocks1}
\bm\alpha_0 &= \bm\Upsilon_{{\rm V}_0{\rm V}_0}  - \bm\Upsilon_{{\rm V}_0{\rm W}_1}\bm\Upsilon_{{\rm W}_1{\rm W}_1}^{-1}\bm\Upsilon_{{\rm W}_1{\rm V}_0},\\
\bm\beta_0 &= \bm\Upsilon_{{\rm V}_0{\rm V}_{1}} - \bm\Upsilon_{{\rm V}_0{\rm W}_1}\bm\Upsilon_{{\rm W}_1{\rm W}_1}^{-1}\bm\Upsilon_{{\rm W}_1{\rm V}_1}, \\ 
\bm\gamma_0 &= \bm\Upsilon_{{\rm V}_1{\rm V}_{0}}  - \bm\Upsilon_{{\rm V}_1{\rm W}_1}\bm\Upsilon_{{\rm W}_1{\rm W}_1}^{-1}\bm\Upsilon_{{\rm W}_1{\rm V}_0}, \\ 
\bm d_0 &= \bm b_{{\rm V}_0}  - \bm\Upsilon_{{\rm V}_0{\rm W}_1}\bm\Upsilon_{{\rm W}_1{\rm W}_1}^{-1}\bm b_{{\rm W}_1},\\
\bm\alpha_{\rm N} &= \bm\Upsilon_{{\rm V}_{\rm N}{\rm V}_{\rm N}}  - \bm\Upsilon_{{\rm V}_{\rm N}{\rm W}_{\rm N}}\bm\Upsilon_{{\rm W}_{\rm N}{\rm W}_{\rm N}}^{-1}\bm\Upsilon_{{\rm W}_{\rm N}{\rm V}_{\rm N}},\\
\bm d_{\rm N} &= \bm b_{{\rm V}_{\rm N}}  - \bm\Upsilon_{{\rm V}_{\rm N}{\rm W}_{\rm N}}\bm\Upsilon_{{\rm W}_{\rm N}{\rm W}_{\rm N}}^{-1}\bm b_{{\rm W}_{\rm N}},
\end{aligned}
\end{equation}
and for $n = 1,\ldots,{\rm N}-1$
\begin{equation}
\begin{aligned}\label{eq:blocks2}
\bm\alpha_n &= \bm\Upsilon_{{\rm V}_n{\rm V}_n} - \bm\Upsilon_{{\rm V}_n{\rm W}_{n}}\bm\Upsilon_{{\rm W}_{n}{\rm W}_{n}}^{-1}\bm\Upsilon_{{\rm W}_{n}{\rm V}_n}  - \bm\Upsilon_{{\rm V}_n{\rm W}_{n+1}}\bm\Upsilon_{{\rm W}_{n+1}{\rm W}_{n+1}}^{-1}\bm\Upsilon_{{\rm W}_{n+1}{\rm V}_n},\\
\bm\beta_n &= \bm\Upsilon_{{\rm V}_n{\rm V}_{n+1}} - \bm\Upsilon_{{\rm V}_n{\rm W}_{n+1}}\bm\Upsilon_{{\rm W}_{n+1}{\rm W}_{n+1}}^{-1}\bm\Upsilon_{{\rm W}_{n+1}{\rm V}_{n+1}},\\
\bm\gamma_n &= \bm\Upsilon_{{\rm V}_{n+1}{\rm V}_{n}} - \bm\Upsilon_{{\rm V}_{n+1}{\rm W}_{n+1}}\bm\Upsilon_{{\rm W}_{n+1}{\rm W}_{n+1}}^{-1}\bm\Upsilon_{{\rm W}_{n+1}{\rm V}_{n}},\\
\bm d_n &= \bm b_{{\rm V}_n} - \bm\Upsilon_{{\rm V}_n{\rm W}_{n}}\bm\Upsilon_{{\rm W}_{n}{\rm W}_{n}}^{-1}\bm b_{{\rm W}_{n}}  - \bm\Upsilon_{{\rm V}_n{\rm W}_{n+1}}\bm\Upsilon_{{\rm W}_{n+1}{\rm W}_{n+1}}^{-1}\bm b_{{\rm W}_{n+1}}.
\end{aligned}
\end{equation}
The block tridiagonal system is never formed in practice, but rather solved on-the-fly using Thomas method (forward Gaussian elimination for tridiagonal matrices). The first equation in \eqref{eq:tridiagonal} is recast as 
\begin{equation}\label{eq:v0}
{\bf V}_0 = \bm\alpha_0^{-1}\lp{\bm d}_0 - \bm\beta_0 {\bf V}_1\rp = \widetilde{\bm d}_0 + \widetilde{\bm\alpha}_0{\bf V}_1\;.
\end{equation}
Using this relation, the second equation reads
\begin{equation}\label{eq:v1}
{\bf V}_1 = \lp\bm\alpha_1 + \bm\gamma_0\widetilde{\bm\alpha}_0\rp^{-1}\lp{\bm d}_1 -\bm\gamma_0\widetilde{\bm d}_0- \bm\beta_1 {\bf V}_2\rp = \widetilde{\bm d}_1 + \widetilde{\bm\alpha}_1{\bf V}_2\;,
\end{equation}
and we can thus establish an analogous expression for the subsequent equations $n = 2,\ldots, {\rm N}-1$, that is
\begin{equation}\label{eq:vn}
{\bf V}_n = \lp\bm\alpha_n + \bm\gamma_{n-1}\widetilde{\bm\alpha}_{n-1}\rp^{-1}\lp{\bm d}_n -\bm\gamma_{n-1}\widetilde{\bm d}_{n-1}- \bm\beta_n {\bf V}_{n+1}\rp = \widetilde{\bm d}_{n} + \widetilde{\bm\alpha}_{n}{\bf V}_{n+1}\;.
\end{equation}
Hence, from the last equation we can retrieve the value for $\bV_{\rm N}$ as
\begin{equation}\label{eq:vN}
{\bf V}_{\rm N} = \lp\bm\alpha_{\rm N} + \bm\gamma_{\rm N -1}\widetilde{\bm\alpha}_{\rm N -1}\rp^{-1}\lp{\bm d}_{\rm N} - \bm\gamma_{\rm N -1}\widetilde{\bm d}_{\rm N -1} \rp  = \widetilde{\bm \alpha}_{\rm N} \widetilde{\bm d}_{\rm N}\;,
\end{equation}
and we then march backwards to recover the remaining $\lbrace \bV_{n}\rbrace_{n={\rm N}-1}^0$ leveraging \eqref{eq:vn}--\eqref{eq:v0}.

\begin{figure}[h!]
 \centering
 \includegraphics[scale = .82]{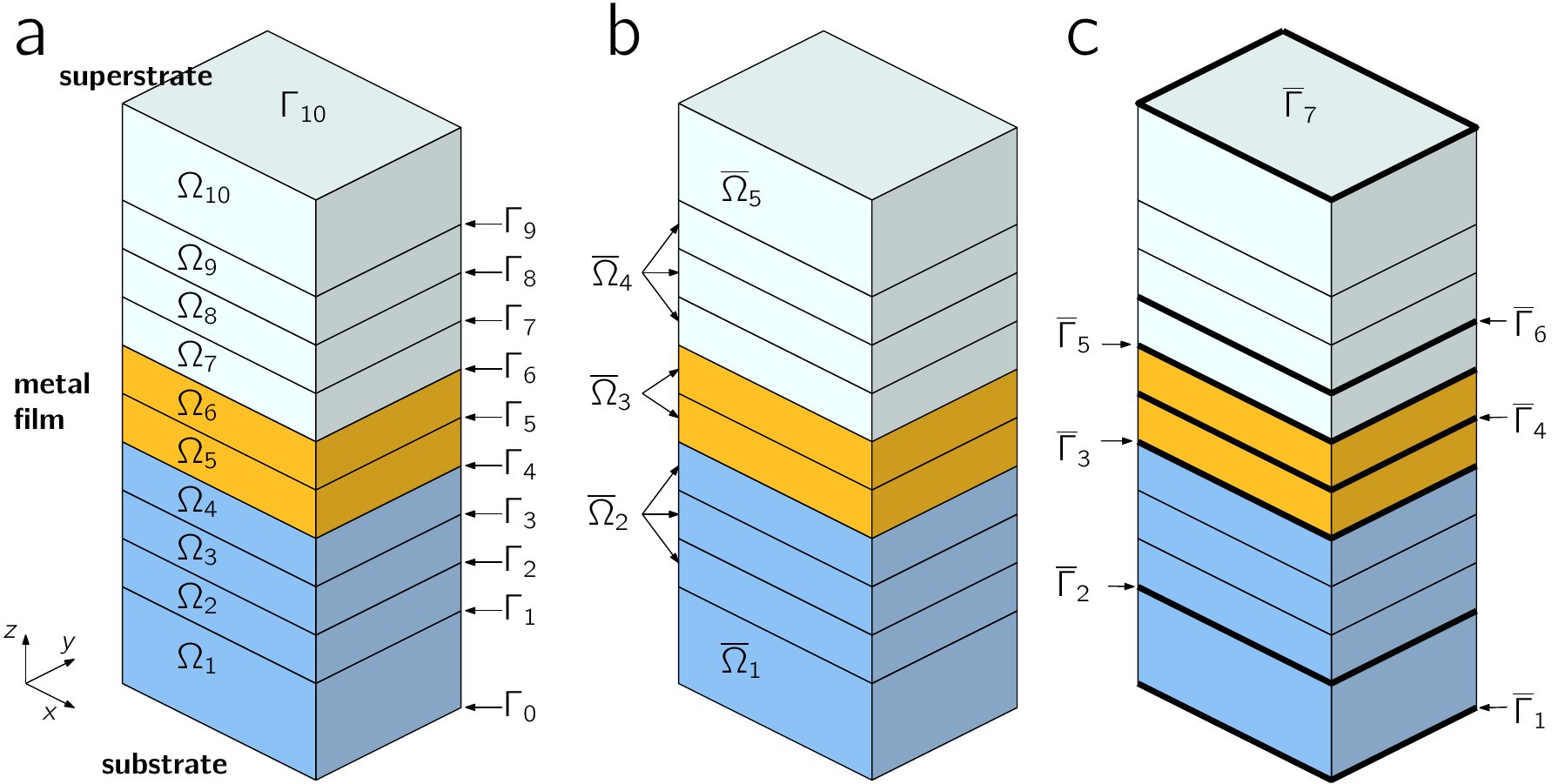}
 \caption{Substrate-metal-superstrate sample structure. (a) Block and interface definition. (b) Block types. (c) Interface types.}\label{fig:types}
\end{figure}

\begin{table}[h!]
\footnotesize
\begin{center}
  \renewcommand{\arraystretch}{1.2}
\begin{tabular}{ccc:cccc}
  \multicolumn{3}{c}{Blocks} & \multicolumn{3}{c}{Interfaces} \\
$\Omega_n$ & $\xrightarrow{\sigma_\Omega}$&$\overline{\Omega}_\ell$ & $\Gamma_n$ & $\xrightarrow{\sigma_\Gamma}$&$\overline{\Gamma}_m$ \\
\hline
1 && 1 & 0&& 1\\
2-4 && 2 & 1-3 &&2 \\
5-6 && 3 & 4&& 3 \\
7-9 & &4 & 5&&4\\
10 & &5 & 6 && 5 \\
 & & &7-9 && 6 \\
 & &  &10 && 7
\end{tabular}
\end{center}
\caption{Block and interface type assignment for structure in Fig. \ref{fig:types}.}\label{tab:types}
\end{table}

\subsection{Implementation}

In this section, we describe how to efficiently compute \eqref{eq:blocks1}-\eqref{eq:blocks2} and solve \eqref{eq:tridiagonal}. The first step is to judiciously partition the computational domain in the extrusion direction to define the blocks and interfaces. For a given block $\Omega_n$, the values on $\bm\Upsilon_{{\rm W}_{n}{\rm W}_{n}}$ depend solely on the spatial dimensions of the block and the material properties ($\varepsilon_{\rm d}$ for dielectric and $\varepsilon_\infty,\,\omega_{\rm p},\,\gamma,\,\beta$ for metal) and for a given interface $\Gamma_n$ the values $\bm\Upsilon_{{\rm V}_{n}{\rm V}_{n}}$ depend solely on the spatial dimensions of the interface and the adjacent blocks' material properties. Hence, accounting for these features when partitioning the domain allows us to define unique block and interface types $\lbrace \overline{\Omega}_\ell\rbrace_{\ell = 1}^L,\;\lbrace \overline{\Gamma}_m\rbrace_{m = 1}^M$ that may translate into important computational savings, similar to the strategy developed by Huynh \etal{} in the context of efficient model order reduction for structured problems \cite{huynh2013static,phuong2013static,eftang2013port,vidal2019multiscale}. The type assignments can be mathematically expressed with a tuple of maps $(\sigma_\Omega,\sigma_\Gamma)$, such that $\sigma_\Omega(n) = \ell$ if block $n$ belongs to type $\ell$, and analogously for the interfaces with $\sigma_\Gamma$. We have illustrated the type definition on a substrate-metal-superstrate structure shown in Fig. \ref{fig:types}, where the block partitioning is done along the $z$ axis. After discretization, we are left with 4, 2 and 4 substrate, metal and superstrate blocks respectively, along with 11 interfaces. Based on the dimensions and the material properties, the type assignment is summarized in Table \ref{tab:types}.

\begin{algorithm}[h!]
\caption{Nested hybridization with full assembly to solve \eqref{eq:globalsystem}}
\label{ag:nhdg}
\begin{algorithmic}[1]
\Require HDG matrix and forcing term $\bm\Upsilon,\,{\bm b}$ of entire domain $\Omega$; type maps $(\sigma_\Omega,\sigma_\Gamma)$ of blocks $\lbrace \Omega_n\rbrace_{n=1}^{\rm N}$ and interfaces $\lbrace \Gamma_n\rbrace_{n=0}^{\rm N}$; six empty lists $v_1,\ldots,v_6$
\For{ $n=1:{\rm N}$}
\State Set $\ell = \sigma_{\Omega}(n)$, $m_0 = \sigma_{\Gamma}(n-1)$ and $m_1 = \sigma_{\Gamma}(n)$ \label{sys0}
\If {$v_1(\ell) =\varnothing$}
\State Compute LU decomposition of $\bm\Upsilon_{{\rm W}_{\ell}{\rm W}_{\ell}} = LU$ and store $LU \mapsto v_1(\ell)$ \label{ag_lu}
\State Solve linear system for multiple right-hand-sides in parallel  \label{sys1}
$$
v_1(\ell) [ \bm x_0,\bm x_1,\bm y] = \lb \bm\Upsilon_{{\rm W}_{\ell}, {\rm V}_{m_0}} ,\, \bm\Upsilon_{{\rm W}_{\ell}, {\rm V}_{m_1}},\, {\bm b}_{{\rm W}_n} \rb,
$$
\hspace{\algorithmicindent}\hspace{\algorithmicindent}\hspace{\algorithmicindent}store $\bm x_0 \mapsto v_2[\ell, m_0],\,\bm x_1 \mapsto v_3[\ell, m_1]$ and set  $\bm x = [\bm x_0,\,\bm x_1,\bm y]$
\Else
\If{$v_2(\ell,m_0) = \varnothing$ {\bf and} $v_3(\ell,m_1) \neq \varnothing$}
\State Solve linear system for multiple right-hand-sides in parallel 
\begin{equation*}
v_1(\ell) [\bm  x_0,\bm y] = \lb \bm\Upsilon_{{\rm W}_{\ell}, {\rm V}_{m_0}} ,\, {\bm b}_{{\rm W}_n} \rb,
\end{equation*}
\hspace{\algorithmicindent}\hspace{\algorithmicindent}\hspace{\algorithmicindent}store $\bm x_0 \mapsto v_2[\ell, m_0]$  and set $x = [\bm x_0,\,v_3[\ell, m_1],\,\bm y]$ \label{sys2}
\ElsIf{$v_2(\ell,m_0) \neq \varnothing$  {\bf and} $v_3(\ell,m_1) = \varnothing$}
\State Solve linear system for multiple right-hand-sides in parallel \label{sys3}
$$
v_1(\ell) [ \bm x_1,\bm y] = \lb \bm\Upsilon_{{\rm W}_{\ell}, {\rm V}_{m_1}} ,\, {\bm b}_{{\rm W}_n} \rb,$$
\hspace{\algorithmicindent}\hspace{\algorithmicindent}\hspace{\algorithmicindent}store $\bm  x_1 \mapsto v_3[\ell, m_1]$ and set $\bm x = [v_2[\ell, m_0],\,\bm x_1,\,\bm y]$
\Else 
\State Solve linear system $v_1(\ell)b= {\bm b}_{{\rm W}_n}$ and set $\bm x = [v_2[\ell, m_0],\,v_3[\ell, m_1],\,\bm y]$ \label{sys4}
\EndIf
\EndIf \label{op2}
\State Multiply and store $\bm\Upsilon_{ {\rm V}_{m_0},{\rm W}_{\ell}} \bm x\mapsto v_4[\ell, m_0],\,\bm\Upsilon_{ {\rm V}_{m_1},\,{\rm W}_{\ell}} \bm x \mapsto v_5[\ell, m_1]$ \label{ag_store}
\State Compute $\bm\alpha_{n-1},\,\bm\beta_{n-1},\,\bm\gamma_{n-1}$ and $\bm d_{n-1}$ using \eqref{eq:blocks1}-\eqref{eq:blocks2} \label{fac1}
\State Compute and store $\widetilde{\bm\alpha}_{n-1}  \mapsto v_6[n,1]$ and $\widetilde{\bm d}_{n-1} \mapsto v_6[n,2]$ using \eqref{eq:v0}-\eqref{eq:vn} \label{fac2}
\If { $n = {\rm N}$}
\State Compute $\bm\alpha_{\rm N},\,\bm d_{\rm N}$ using \eqref{eq:blocks1} and $\widetilde{\bm\alpha}_{\rm N},\,\widetilde{\bm d}_{\rm N}$ using \eqref{eq:vN}
\State Evaluate ${\bV}_{\rm N} = \widetilde{\bm\alpha}_{\rm N}\widetilde{\bm d}_{\rm N}$ \label{fac3}
\EndIf
\EndFor
\For {$ n = {\rm N}:-1:1$  {\bf with} $\ell = \sigma_{\Omega}(n)$,  $m_0 = \sigma_{\Gamma}(n-1)$, $m_1 = \sigma_{\Gamma}(n)$} \label{op3} 
\begin{align*}
{\bV}_{n-1} &= v_6(n,2) + v_6(n,1){\bV}_{n}\\
v_1(\ell){\bW}_{n} &= {\bm b}_{{\rm W}_n} - \bm\Upsilon_{{\rm W}_\ell{\rm V}_{m_0}} {\bV}_{n-1} - \bm\Upsilon_{{\rm W}_\ell{\rm V}_{m_1}} {\bV}_{n}
\end{align*}
\EndFor
\end{algorithmic}
\end{algorithm}

Once the type maps have been established and the matrix and forcing term $\bm\Upsilon,\,\bm{b}$ have been computed, the degrees of freedom $\widehat{\bU}$ following Algorithm \ref{ag:nhdg}, a procedure that we refer to as nested HDG with full assembly. Since an integral part of this nested hybridization method is the ability to reuse computations by virtue of the block and interface types, we shall define six lists $\lbrace v_i\rbrace_{i=1}^6$ where all relevant computations are stored and can thus be accessed whenever required by Algorithm \ref{ag:nhdg}. These lists will contain the following items: $v_1$ stores the LU decomposition of $\bm\Upsilon_{{\rm W}{\rm W}}$ for each block type, and are indexed by block type; $v_2,\ldots,v_5$ contain the building blocks of the Schur decomposition at the block-interface level, and are indexed by both block and interface type, see Algorithm \ref{ag:nhdg} for the exact expressions; finally, $v_6$ contains each intermediate factor in the solution of the tridiagonal system \eqref{eq:tridiagonal} using forward block Gaussian elimination, that is $\lbrace \widetilde{\bm\alpha}_n,\widetilde{\bm d}_n\rbrace_{n=0}^{\rm N}$, which are required to compute the degrees of freedom for $\bV$ with equations \eqref{eq:v0}--\eqref{eq:vn}.

We make the following remarks regarding Algorithm \ref{ag:nhdg}: (i) instead of precomputing the HDG matrix $\bm\Upsilon$ and forcing ${\bm b}$, which may require significant storage, the matrix and forcing term can be partially assembled on-the-fly after operation \ref{sys0}, that is only for the degrees of freedom $\lbrace {\rm W}_{\ell},{\rm V}_{m_0},{\rm V}_{m_1}\rbrace$ and purged after each iteration, a variation that is referred to hereafter as nested HDG with partial assembly; however, if this strategy is pursued both matrix and forcing term need to be partially re-assembled in operation \ref{op3} to recover the solution field, hence the saving in memory (the HDG matrix is never entirely assembled) comes at the expense of a higher computational runtime; (ii) since storing and reusing computations is an integral part of the algorithm, we may also eliminate items from $v_2,\ldots,v_5$ as soon as they are no longer needed; (iii) the solution of the linear systems for multiple right-hand-sides in steps \ref{sys1}, \ref{sys2} and \ref{sys3} of the algorithm is the most computationally intensive, although it can be trivially parallelized; (iv) the storage of $v_6$ requires significant memory storage since all steps are needed to recover the interface degrees of freedom as in operation \ref{op3}; and (v) this hybridization results in a strong compression of the original problem, since once $\lbrace{\bV}_n\rbrace_{n=0}^{\rm N}$ have been recovered we may evaluate $\lbrace{\bW}_n\rbrace_{n=1}^{\rm N}$ through operation \ref{op3} block-wise at minimal cost (the LU decompositions are already stored in $v_1$) and then obtain the local variables $\lbrace \bE_h,\,\bH_h,\,\bJ_h,\,\uprho_h\rbrace$ with the classical HDG static condensation expressions at the element level, see \cite{vidal2017simulation,vidal2018hybridizable}.

The computational strategy described above is a purely algebraic construction, hence it is not only applicable to the HDG discretization of Maxwell's equations augmented with the hydrodynamic model, but to any linear system arising from an HDG discretization. In order to minimize the computational and memory requirements that stem from the nested HDG method described in Algorithm \ref{ag:nhdg}, a carefully designed mesh and judicious degree-of-freedom choice is critical.


\section{Numerical results}\label{sec:res}
The numerical results presented in this section have been simulated with the \texttt{MATLAB} implementation of the nested HDG method described above and the classical HDG for Maxwell's equations introduced in \cite{nguyen2011maxwell,vidal2018hybridizable}. The computational times and memory footprint of the simulations correspond to a 512GB Linux 18.04 machine with 16 AMD Opteron(tm) Processors 6320x15 that has been used to perform the simulations.

\subsection{Plane wave through layered media}
\revtwo{In this section, we perform a numerical test to verify the implementation and accuracy of the nested HDG. To that end, we use a plane wave propagating through a sapphire-silica-air layered medium under normal incidence, for which the exact solution is known, and compare the errors of classical HDG, \ie direct solution of \eqref{eq:globalsystem}, with nested hybridization described by Algorithm \ref{ag:nhdg}. }

\begin{table}[h!]
\footnotesize
\begin{center}
  \renewcommand{\arraystretch}{1.2}
\begin{tabular}{cc|cccc|cccc}
&  & \multicolumn{4}{c}{HDG} & \multicolumn{4}{c}{nested HDG} \\
 &   & \multicolumn{2}{c}{$\norm{\bE_0- \bE_h}_{\bm L^2}$} & \multicolumn{2}{c}{$\norm{\bE_0- \bE_h}_{\bm H^{\curl}}$} &\multicolumn{2}{c}{$\norm{\bE_0- \bE_h}_{\bm L^2}$} &\multicolumn{2}{c}{$\norm{\bE_0- \bE_h}_{\bm H^{\curl}}$}\\
$p$& $n$  & Error& Order& Error& Order& Error& Order & Error& Order \\
[1mm]
2 & 8 & 5.0e-2 & --   & 2.2e-1 & --   & 5.0e-2 & --   & 2.2e-1 & --   \\
  &16 & 2.6e-3 & 4.27 & 4.7e-2 & 2.20 & 2.6e-3 & 4.27 & 4.7e-2 & 2.20 \\
  &32 & 1.2e-4 & 4.42 & 1.2e-2 & 2.01 & 1.2e-4 & 4.42 & 1.2e-2 & 2.01  \\
  &64 & 9.7e-6 & 3.63 & 2.9e-3 & 2.00 & 9.7e-6 & 3.63 & 2.9e-3 & 2.00   \\
  [2mm]
3 & 8 & 2.9e-3 & --   & 4.5e-2 & --   & 2.9e-3 & --   & 4.5e-2 & --   \\
  &16 & 6.6e-5 & 5.47 & 5.6e-3 & 3.01 & 6.6e-5 & 5.47 & 5.6e-3 & 3.01\\
  &32 & 3.0e-6 & 4.46 & 7.2e-4 & 2.97 & 3.0e-6 & 4.46 & 7.2e-4 & 2.97  \\
  &64 & 1.7e-7 & 4.14 & 9.0e-5 & 2.99 & 1.7e-7 & 4.14 & 9.0e-5 & 2.99  \\
  [2mm]
  4 & 8 & 1.6e-4 & --   & 6.6e-3 & --   & 1.6e-4 & --   & 6.6e-3 & --   \\
  &16 & 3.5e-6 & 5.51 & 4.8e-4 & 3.79 & 3.5e-6 & 5.51 & 4.8e-4 & 3.79  \\
  &32 & 9.3e-8 & 5.23 & 3.1e-5 & 3.96 & 9.3e-8 & 5.23 & 3.1e-5 & 3.96  \\
  &64 & 2.8e-9 & 5.08 & 1.9e-6 & 3.99 &  2.8e-9 & 5.08 & 1.9e-6 & 3.99
  \end{tabular}
\end{center}
\caption{History of convergence for the HDG and nested HDG solution.} \label{tab:hdghydro}
\end{table}
\revtwo{The refractive indices of the layers are 3.31 (sapphire), 1.98 (silica) and 1 (air), with respective thicknesses of 500, 250 and 250 nm. The plane wave is an $x$-polarized 1 micron wavelength plane wave propagating in the positive $z$-direction, impinging from the sapphire layer. The computational domain is a prism of $250\times250\times1000$ nm discretized in $6\times 6\times n$ isotropic cubes, and we prescribe $\bE\times\bn = \bm 0$ on the $x$-constant  boundaries and $\bH\times\bn = \bm 0$ on the $y$-constant boundaries, as well as first-order absorbing conditions on the top and bottom boundaries. We focus on $p=2,\,3,\,4$ and several $n$ values, and compute the $\bm L^2(\mcT{})$ and $\bm H^{\rm curl}(\mcT{})$ errors of $\bE_h$, collected in Table \ref{tab:hdghydro} along with a convergence analysis. Note that since the cubes in the $z$-direction are isotropic, there are only three different types of blocks (sapphire, silica, air) and seven interfaces (upper-lower boundaries, sapphire-sapphire, sapphire-silica, silica-silica, silica-air and air-air).}


\begin{figure}[h!]
 \centering
 \includegraphics[scale = .73]{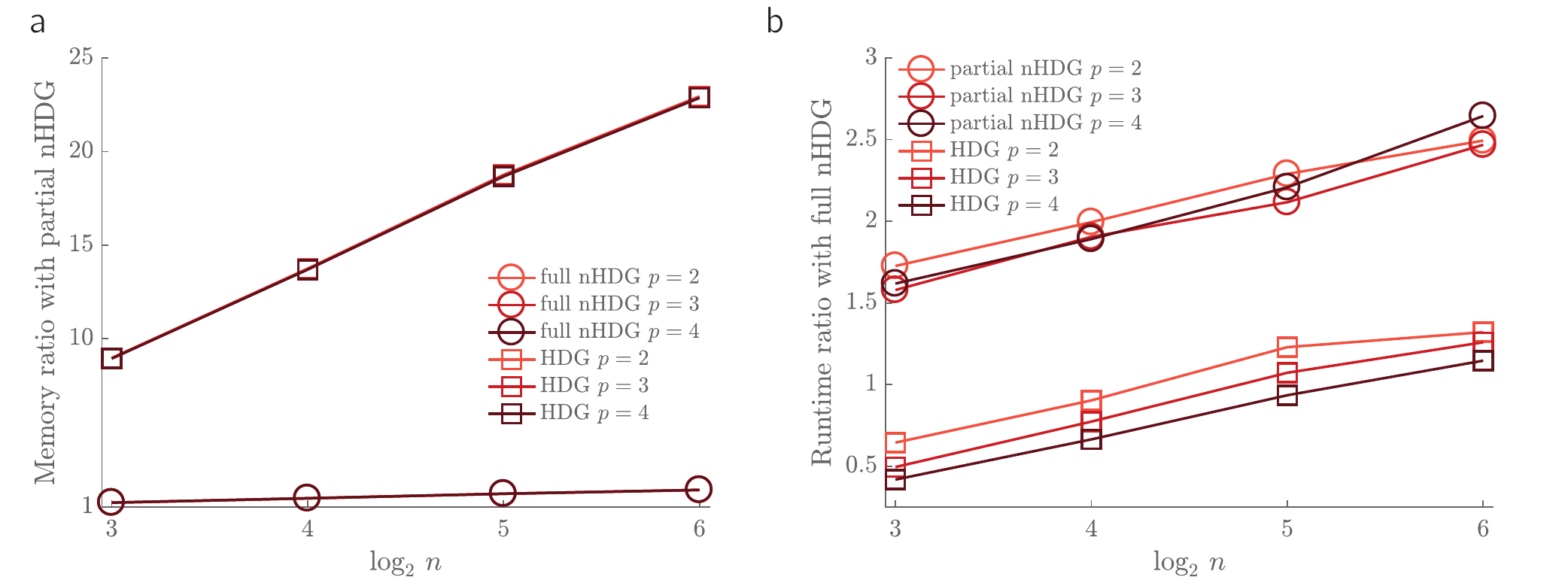}
 \caption{Comparison of computational costs between methods, for $p=2,\,3,\,4$ and $n = 8,\,16,\,32,\,64$: (a) Memory requirements with respect to nested HDG with partial assembly. (b) Runtime with respect to nested HDG with full assembly.}\label{fig:compare_cost}
\end{figure}

\revtwo{As expected, the solutions computed by both methods have the exact same errors and orders of convergence, since the nested HDG is just an algebraic modification that enables a more efficient solution of the classical HDG linear system. We now compare the computational costs and memory requirements of HDG to the nested version, both with full and partial assembly. The computational runtimes correspond to the wall time averages of 10 individual simulations for each method; the memory is measured in terms of RAM GB required to execute the algorithms. The most memory-efficient method is the nested HDG with partial assembly, and the differences become starker as the mesh is refined, requiring half the memory of that of nested HDG with full assembly and about 20 times less than HDG for the finest mesh, see Fig. \ref{fig:compare_cost}(a). In terms of computational runtime, the nested HDG with full assembly, which we use to benchmark in Fig. \ref{fig:compare_cost}(b), is obviously faster than nested HDG with partial assembly due to the cost of operation \ref{op3}. Even though solutions computed with classical HDG are faster for coarser meshes, as the discretization is refined our implementation of nested HDG with full assembly becomes faster due to the reuse of computations at the block level. The main takeaway from this example is that the nested HDG produces the same solutions as classical HDG while exhibiting significantly lower memory requirements; the differences in computational runtime depend on a myriad of factors, namely the resolution of the 2-D mesh (the interfaces), the amount of unique blocks, the number of processors available and the efficiency of the implementation, to name a few. However, based on the results in this article, we can conclude that the nested HDG has the potential of resolving the linear HDG system faster than via direct solution.}


\subsection{Triangular nanocoaxial aperture}
We now consider a metallic nanostructure that produces extraordinary optical transmission and can excite second harmonic fields. This structure consists of periodic arrays of subwavelength triangular apertures of a dielectric material patterned in a metallic film, and unlike arrays of annular nanogap structures that have been simulated with HDG in previous works   \cite{park2015nanogap,yoo2016high,vidal2018hybridizable}, triangular apertures are not centrosymmetric, a requirement to excite second-order effects. 

\subsubsection{Structure definition}
\begin{figure}[h!]
 \centering
 \includegraphics[scale = .6]{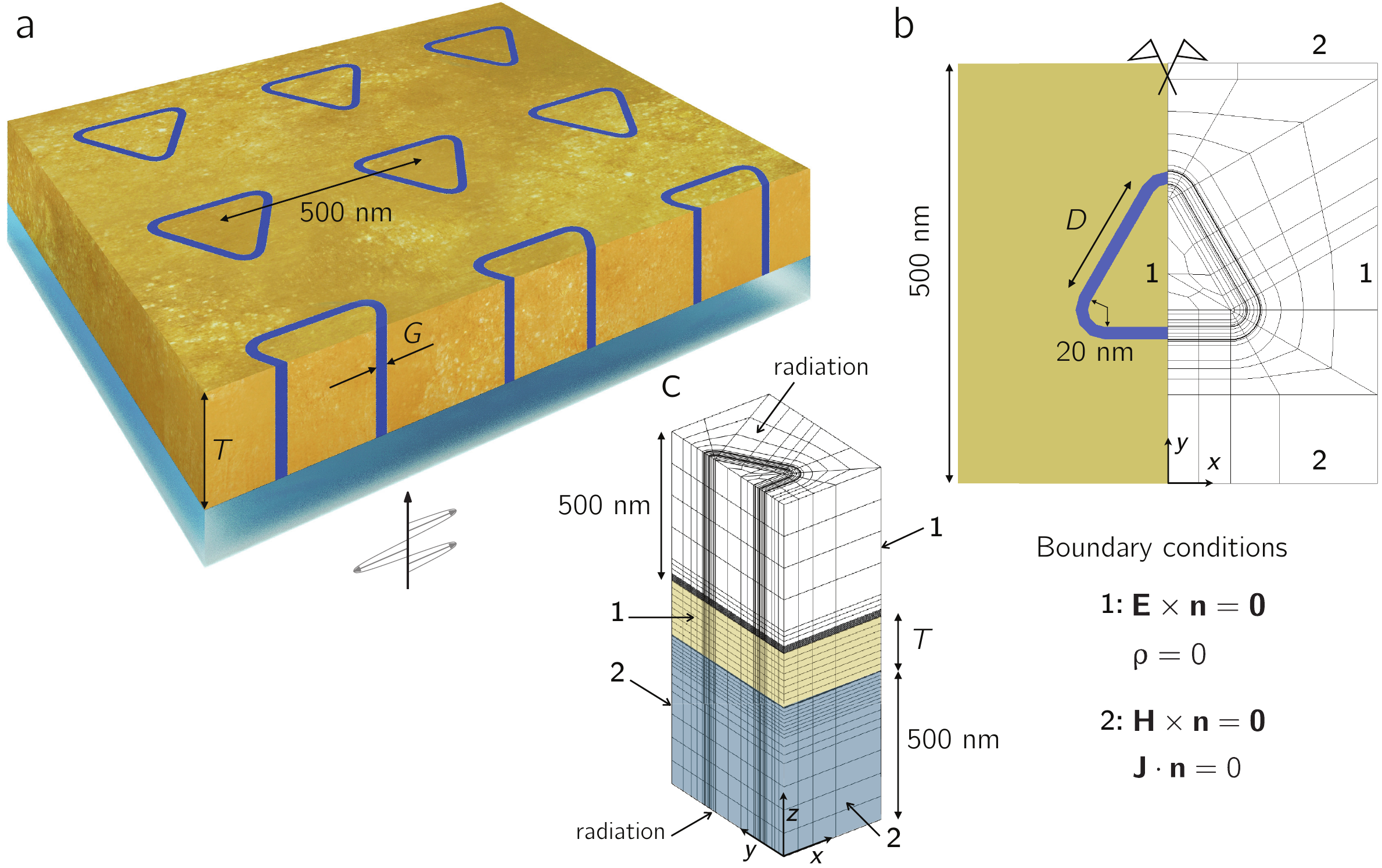}
 \caption{Triangular coaxial nanostructure: (a) 3-D model of periodic array with relevant measures. (b) Cross-sectional view and 2-D curved mesh of unit cell, with symmetry axis and boundary condition specification. (c) 3-D computational domain of unit cell.}\label{fig:triangle}
\end{figure}


The structure that will be analyzed is a gold thin-film with triangular coaxial nanogaps arranged according to the symmetries of the square, see Fig. \ref{fig:triangle}(a), for wavelengths ranging from visible to low infra-red. The metal film is deposited over a sapphire substrate, a transparent material in these frequency regimes, and the nanogap is filled with alumina. The structure is illuminated from below with an $x$-polarized plane wave, and we can exploit the symmetries of the structure and solve only for the domain shown in Fig. \ref{fig:triangle} containing half of the triangular nanogap. Under these symmetry conditions, we prescribe $\bE\times\bn = \bm 0,\;\uprho = 0$ on the $x$-constant  boundaries and $\bH\times\bn = \bm 0,\;\bJ\cdot\bn = 0$ on the $y$-constant  boundaries. First-order radiation conditions are imposed on the $z$-constant boundaries. To assess the efficacy of SHG, we monitor the transmittance $\pow$ through the structure and the second harmonic transmittance, computed as
\begin{equation}
  \pow_1 = 100\cdot\frac{\abs{\int_{A_o} \Re \lb \bE_1 \times \bH_1^*\rb \cdot \bn \ {\rm d}{ A}}}{\abs{\int_{A_i} \Re \lb \bE_{\rm inc} \times \bH_{\rm inc}^*\rb \cdot \bn \ {\rm d}{ A}}}\,,\qquad \qquad
\pow_2 = 100\cdot\frac{\abs{\int_{A_o} \Re \lb \bE_2 \times \bH_2^*\rb \cdot \bn \ {\rm d}{ A}}}{\abs{\int_{A_i} \Re \lb \bE_{\rm inc} \times \bH_{\rm inc}^*\rb \cdot \bn \ {\rm d}{ A}}}\,, 
\end{equation}
where $A_i$ is an arbitrary $xy$ plane below the gold film and $A_o$ an arbitrary $xy$ plane above the gold film. 
 
 \begin{figure}[h!]
 \centering
 \includegraphics[scale = 1]{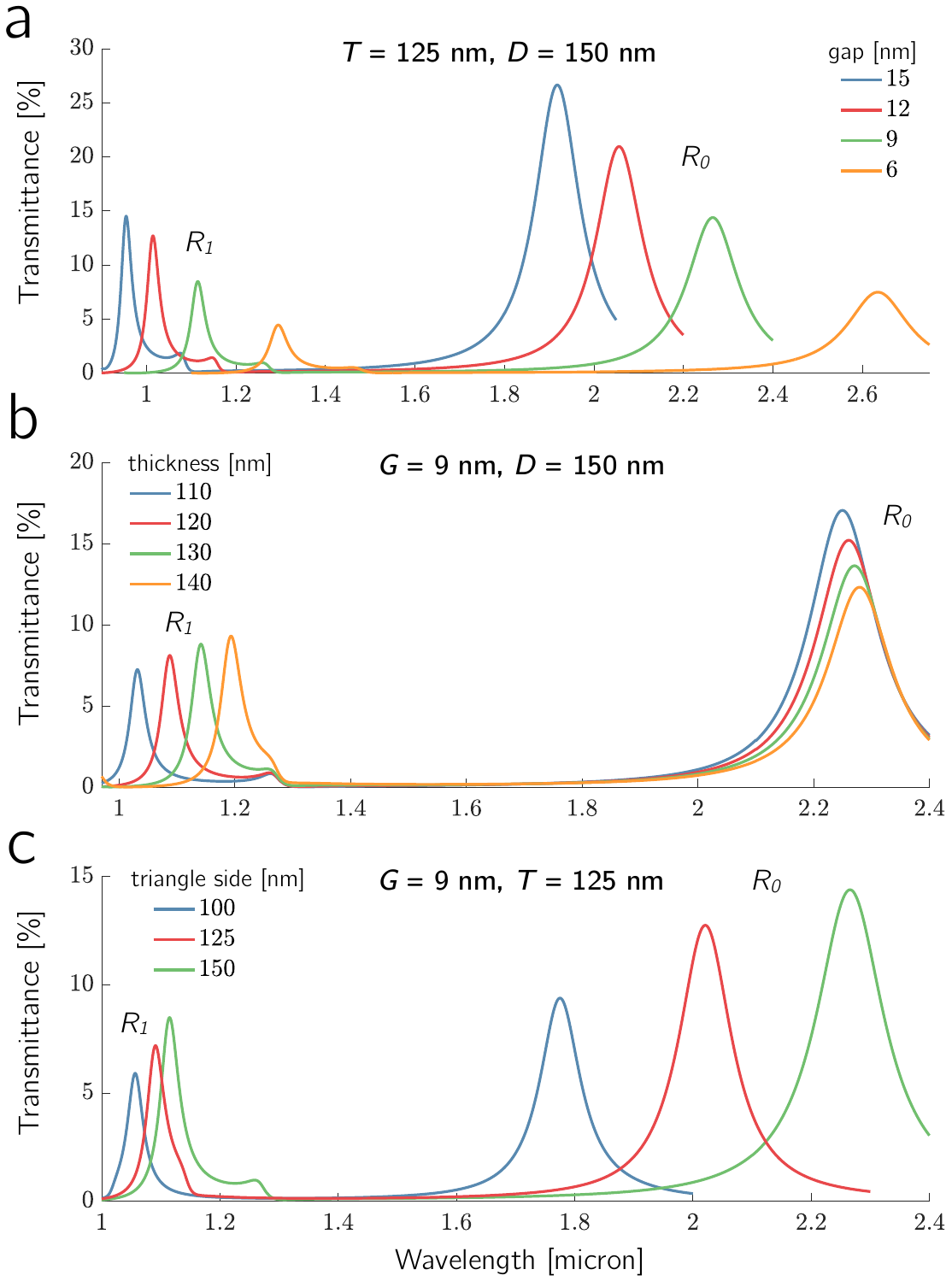}
 \caption{Wavelength [$\upmu$m] - transmittance [\%] curves (x-axes shared between subfigures): (a) Effect of decreasing gap. (b) Effect of increasing thickness. (c) Effect of increasing triangle side length.}\label{fig:geometryimpact}
\end{figure}

The discretization consists of 18K hexahedral cubic elements, and is constructed by extruding in the $z$-direction the 2-D curved mesh in Fig. \ref{fig:triangle}(b). The 2-D curved mesh, with 330 elements, is devised such that the rounded corners are properly represented, and a boundary-layer type discretization is used for the region surrounding the gap. To that end, we place 2-D coaxial layers at distances 0.5, 1, 2, 3 and 5 nm on both sides of the gap-metal interfaces, ensuring enough resolution for both first and second order phenomena.  In the vertical direction, we set both the substrate and superstrate thickness to 500 nm which is sufficient to properly represent illumination conditions and domain unboundedness. For each stratum, we use 19 blocks divided among 4 types, increasing the thickness of each block type as we move further away from the gold film. This computational strategy allows us to capture  the rapidly-varying near-field effects in the vicinity of the metal surface (< 5 nm) as well as to smoothly transition towards the far-field values of transmittance. For the gold stratum, we use one thin block type for the regions near the metal-substrate and metal-superstrate interfaces (4 blocks of 0.25 nm thickness for both upper and lower areas) to capture the boundary-layer features that develop and one thick block type for the rest (8 blocks). Hence, the total number of blocks is 54, split into 10 type blocks and 7 type interfaces, see Fig. \ref{fig:triangle}(c). The mesh topology we have described is the same for the different geometric parameters discussed below, and since the hydrodynamic density profile is independent of the gap size we use the same values of 2-D coaxial layers and 3-D boundary-layer detail at the upper and lower metal surfaces for all gaps. Numerical accuracy is verified by carrying out grid convergence studies on consecutively refined meshes, until the relative error for the SHG transmittance is below 1\%. This highly anisotropic mesh, along with the nested HDG method, allows us to efficiently solve for the full 3-D EM wave field.

\begin{figure}[h!]
 \centering
 \includegraphics[scale = .58]{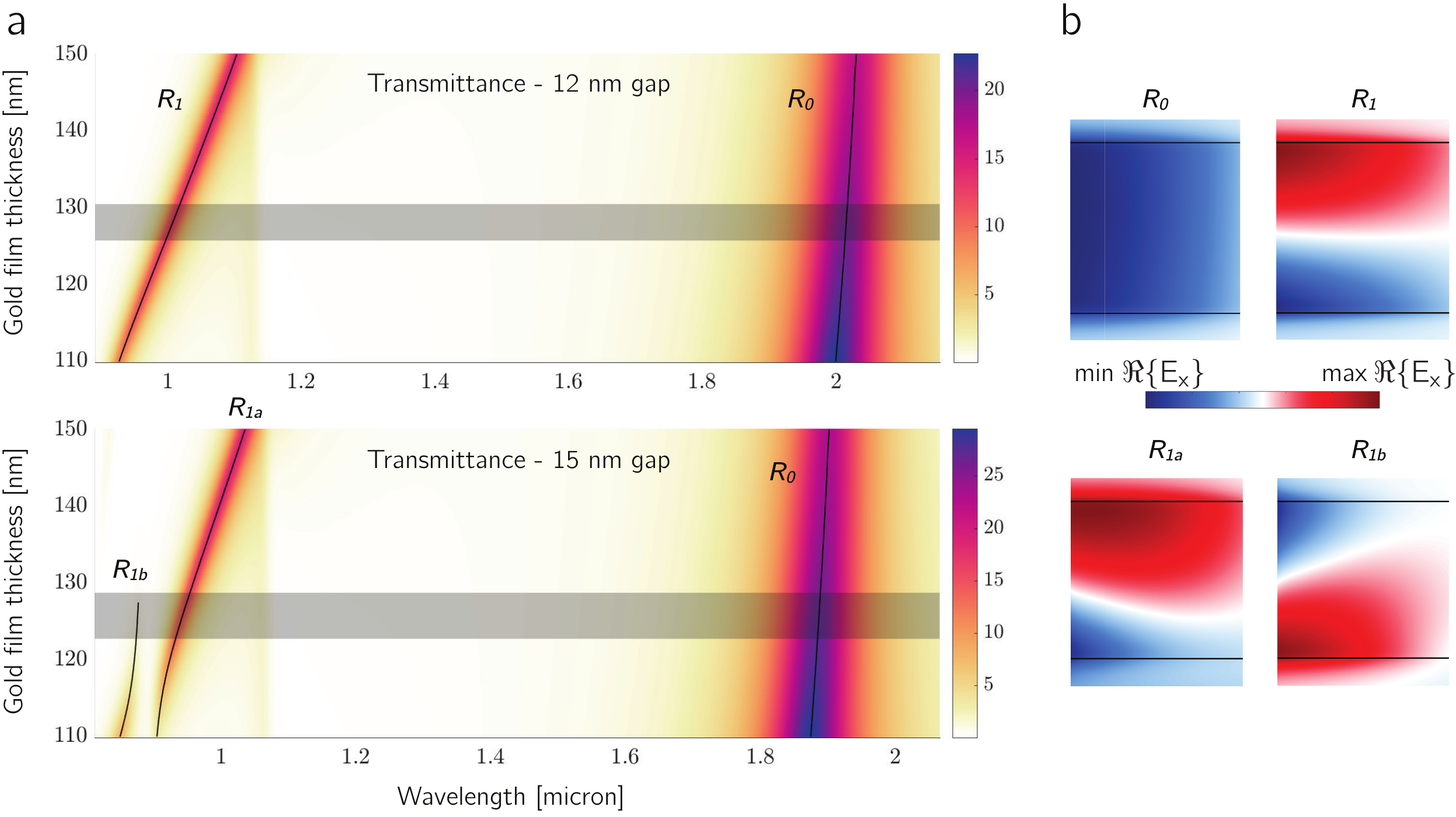}
 \caption{(a) Transmittance [\%]-wavelength [$\upmu$m]-thickness [nm] heatmaps for 12 nm (top) and 15 nm gap (bottom), with lines to visually aid resonance tracking and gray shaded area to identify thicknesses that lead to doubly-resonant structures. (b) Real part of ${\rm E_x}$ for the different resonances along the mid-gap diagonally-vertical plane. Black lines correspond to upper and lower film surfaces.}\label{fig:heatmap}
\end{figure}

\subsubsection{Optimal geometry}\label{sec:rom}

For the structure under consideration, we set the array periodicity to 500 nm and the radius of curvature at the triangle corners to 20 nm for fabrication purposes. The remaining geometric features, namely the gap size $G$, the triangle side $D$ and the film thickness $T$ need to be numerically determined so as to achieve a doubly-resonant structure. The idea behind such a structure is that resonances are excited at both $\omega$ and $2\omega$, thus amplifying the second-order effects that are generated as a consequence of its non-centrosymmetric nature. 

\revone{Firstly, we need to understand the effect of $G,\,D$ and $T$ on the resonances. In order to alleviate the computational burden of this parametric study, we perform the simulations using the Drude model (by setting $\beta = 0 $ in  \eqref{eq:fhg}) instead of the hydrodynamic model; in the mid-infrared the Drude model predicts a transmittance spectrum that is qualitatively identical to that of the hydrodynamic (only red-shifted), hence it suffices to understand the impact of the geometry parameters. Furthermore, the high-resolution mesh defined above is specifically tailored to capture SHG, hence we may use a coarser mesh just for these geometry simulations since the Drude model does not solve for the \AA ngstrom-thin accumulation charge layers at the metal-dielectric interfaces. To that end, we use a mesh similar to that of Fig. \ref{fig:triangle}(b,c), but with only 112 2-D elements for a total of 2K hexahedral cubic elements.}

\revone{For a single wavelength, simulating the full electromagnetic response using the Drude model on the coarse mesh requires solving a sparse linear system of size 190K and 62M non-zeros (1.5GB of RAM), which takes 9 minutes to solve using \texttt{MATLAB}'s backslash operation. We simulate the full spectra of transmittance (0.9-2.8 micron) for several gaps, triangle side lengths and film thicknesses. The impact of gap is shown in Fig. \ref{fig:geometryimpact}(a) for $T=125$ nm and $D = 150$ nm, the impact of thickness is shown in Fig. \ref{fig:geometryimpact}(b) for $G=9$ nm and $D = 150$ nm and the impact of triangle side length is shown in Fig. \ref{fig:geometryimpact}(c) for $T=125$ nm and $G = 9$ nm. The wavelength $\lambda_0^*$ of resonance $R_0$ is blue-shifted as $D$ decreases, whereas it is almost insensitive to the thickness. Conversely, the wavelength $\lambda_1^*$ of resonance $R_1$ is much more sensitive to changes in $T$ (blue-shift for decreasing thickness) than to changes in $D$. Finally, both resonances are similarly affected by gap size modifications. Consequently, for a given gap size one should fix either $D$ (resp. $T$) and vary $T$ (resp. $D$) to attain a geometric configuration that is doubly-resonant at frequencies $\omega$ and $2\omega$.}

\revone{For this structure, we choose to fix $D$ and optimize the double resonance as $T$ varies, and leverage our previous work to compute parametrized solutions of plasmonic structures \cite{vidal2018computing}, whereby a small number of high-fidelity simulations can be used to construct an accurate reduced order model (ROM) that enables the inexpensive computation of approximate solutions. The reduced order model is constructed upon two parameters: the incident wavelength, to obtain spectrum profiles; and the film thickness, to evaluate how variations in the film thickness impact the resonances. To achieve a parametric representation of the thickness, we build a mapping using $\mcal{C}^2$ splines that prescribes deformations in the z-direction, thus ensuring that thickness variations starting from a reference thickness value may be accommodated and are continuous and differentiable. Further details on how to parametrize geometry in plasmonics using deformation mappings may be found in \cite{vidal2018computing}.}

\revone{For each gap size of interest, we set the triangle side to $D=150$ nm and build a ROM by first computing 200 high-fidelity simulations (Drude model on the coarse mesh) and then combining these solutions, or snapshots, to form a low-dimensional approximation space, see \cite{vidal2018computing}. This is commonly known as offline stage, which is computationally intensive (each of the 200 solutions takes 9 minutes for a total of 30 hours) but done only once. After completing this stage, the main advantage of ROMs is that they can be queried for any value of $\lambda \in [0.8,\,22]$ micron and $T \in [110,\, 150]$ nm --these are the prescribed intervals of interest for the triangular coax--  and produce an approximate full-wave 3-D solution of \eqref{eq:fhg} in less than 0.1 seconds. This multi-query process, known as the online stage, will obviously exhibit lower accuracy since instead of the high-fidelity solver we employ a surrogate model. However, for the reduced order models under consideration we report relative errors in transmittance of less than 5\% when comparing the ROM solution to the high-fidelity HDG solution, hence the ROM are a suitable computational tool to study the impact of thickness in the resonances of this triangular coaxial structure.}

{In this case, the burden of the offline stage is greatly compensated by the efficiency of the online stage, since transmittance-wavelength-thickness heatmaps can be obtained by inexpensively querying the ROM for multiple $\lp \lambda,T\rp$ combinations, which would otherwise require a full HDG 3-D simulation for each $\lp \lambda,T\rp$. These heatmaps are paramount to track the resonances as a function of the thickness and to identify, for each gap, the metal film thickness that gives rise to a doubly-resonant structure. We show the transmittance heatmaps for 12 and 15 nm nanogaps with $D = 150$ nm in Fig \ref{fig:heatmap}, where we notice that the $R_1$ resonance splits between two for 15 nm gaps and above, whereas it remains a single resonance for gaps below 15 nm. A field plot of the real part of the $x$-electric amplitude is provided in Fig. \ref{fig:heatmap}, where the field is shown along the mid-gap diagonal plane --that is, the vertical plane that runs along the middle of the gap and is parallel to the longer side of the triangle as shown in Fig. \ref{fig:geometryimpact}(b)-- where it can be observed that the $R_1$ splitting gives rise to modes $R_{1a},\,R_{1b}$ that are not constant along the diagonal direction, as opposed the $R_1$ mode. The shaded gray area corresponds to thickness values that lead to double resonances.}



\begin{figure}[h!]
 \centering
 \includegraphics[scale = .85]{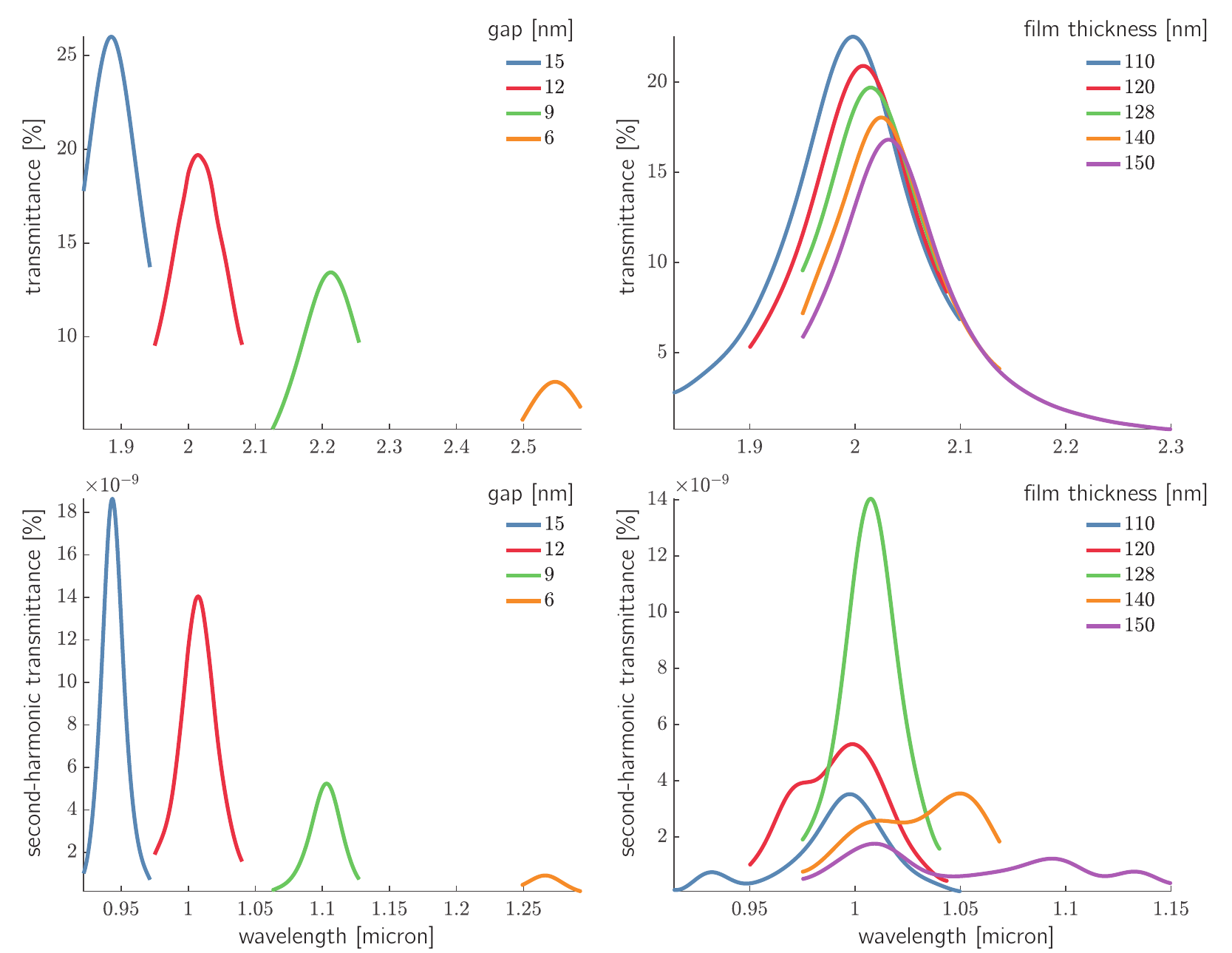}
 \caption{First and second harmonic transmittance for $D=150$ nm. (a,b) Transmittance profile $\pow_1$ (a) and $\pow_2$ (b)-wavelength [$\upmu$m] for 6, 9, 12 and 15 nm gap and optimal thickness to realize double resonances. (c,d) Transmittance profile $\pow_1$ (c) and $\pow_2$ (d)-wavelength [$\upmu$m] for the 12 nm gap and several film thicknesses. Optimal thickness leading to double resonance is 128 nm.}\label{fig:shg_response}
\end{figure}

\subsubsection{Second-harmonic simulations}
\revone{Once we have identified, for each gap, the values of $D$ and $T$ that excite modes at $\omega$ and $2\omega$, we can apply the computational strategy summarized in \eqref{eq:fhg}-\eqref{eq:shg} on the fine mesh and with the hydrodynamic model ($\beta>0$) to compute SHG, where we solve the global HDG linear systems by means of Algorithm \ref{ag:nhdg}. In order to simulate the spectra shown in Fig. \ref{fig:shg_response}, for each gap size and film thickness we solve \eqref{eq:fhg}-\eqref{eq:shg} for 24 different wavelength values. These simulations are expensive, taking around 19 hours --9.5 hours for each \eqref{eq:fhg} and \eqref{eq:shg}-- per wavelength, for a grand total of 19 days of nonstop computation to recover the first and second harmonic spectrum for a given gap and thickness. We now discuss the breakdown of simulation costs of either \eqref{eq:fhg} or \eqref{eq:shg} for one wavelength into operations as per Algorithm \ref{ag:nhdg}, using the nested HDG with partial assembly and purging the lists $v_2,\,v_5$ of unnecessary computations after each iteration. The LU decomposition (operation \ref{ag_lu}) takes 25 min, solving the linear system (operations \ref{sys1}, \ref{sys2}, \ref{sys3}, \ref{sys4}) takes 250 min, forming and operating the full matrices that result from the second hybridization (operations \ref{fac1}, \ref{fac2}, \ref{fac3}) take 282 min and the $\lbrace{\bW},{\bV}\rbrace$ recovery (operation \ref{op3}) takes 3 min. In terms of storage, the partial assembly of $\bm\Upsilon$ requires 0.5 GB, the LU decompositions in $v_1$ requires 13 GB for all types, $v_2$ and $v_3$ combined require a maximum of 29 GB, $v_4$ and $v_5$ combined require a maximum of 28 GB and finally $v_6$ requires 128 GB. The storage requirement at any given algorithm step does not exceed 150 GB thanks to the type blocks and interfaces definition and since the information on $v_2,\ldots,v_5$ can be eliminated as the algorithm progresses. However, the storage for $v_1,\,v_6$ keeps increasing throughout the main loop, and it can only be deleted after it has been used to recover $\lbrace{\bW},{\bV}\rbrace$  in operation \ref{op3}. Unfortunately, no specific cost comparison can be drawn with classical HDG because the direct solution of the fully assembled sparse matrix exceeds the RAM capacity of our machine (512 GB). If fully assembled, the HDG linear system $\bm\Upsilon$ is of dimension 2M, with 750M non-zeros for a total of 18 GB in storage.}

\revone{The bottleneck for the nested HDG method is the discretization of the 2-D mesh rather than the amount of blocks in the third dimension. The resolution of the 2-D mesh will dictate the amount of degrees of freedom at the interfaces, which is directly related to the amount of forcing terms when solving the linear systems (operations \ref{sys1}, \ref{sys2}, \ref{sys3}), as well as to the dimension of the matrices $\bm \alpha_n$ that need to be inverted to solve the tridiagonal system, which are no longer sparse. A possible strategy to mitigate the computational costs is to increase the number of processors, since operations \ref{sys1}, \ref{sys2}, \ref{sys3} are embarrassingly parallel with respect to the number of forcing terms. However, further research is needed to reduce the computational burden of the forward Gaussian elimination of the block tridiagonal matrix (Thomas algorithm), summarized in operations \ref{fac1}, \ref{fac2}, \ref{fac3}.}

For all simulations, we choose an illumination intensity of $100\, {\rm MW/cm}^2$, which corresponds to a reference magnetic field $\alpha = 7.29\cdot10^{4}\,{\rm A/m}$, and a reference lengthscale $L_{\rm c} = 10^{-9}$ m. The values for gold optical constants are  $\varepsilon_\infty=1$, $\hbar\widebar{\omega}_{\rm p} = 8.45$ eV, $\hbar\widebar{\gamma} = 0.047$ eV \cite{olmon2012optical}, the Fermi velocity $\widebar{v}_{\rm F} = 1.39\cdot 10^6$ m/s, the equilibrium charge density $\widebar{n}_0 = 5.91 \cdot 10^{28}\,\rm{m}^{-3}$ \cite{ashcroft2005solid} and finally the electron charge $\widebar{e} = 1.062\cdot10^{-19}\,{\rm C}$. The dielectric constant values for alumina as a function of the incident wavelength are given by \cite{boidin2016pulsed} for $\lambda < 1.5$ micron and \cite{kischkat2012mid} for $\lambda > 1.5$ micron, whereas the permittivity of sapphire is taken from \cite{malitson1972refractive}.

The first and second order transmittances $\pow_1,\pow_2$ are shown in Figs. \ref{fig:shg_response}(a) and (b) for 6, 9, 12 and 15 nm gap. Setting $D=150$ nm, we capitalize on the ROMs constructed in Section \ref{sec:rom} and find that thickness values of 128.5, 129, 128 and 126 nm lead to double resonances for these gaps, respectively. Even though the ROM is constructed for the Drude model only, the doubly-resonant thickness is still valid for the nonlocal calculations, since the blue-shift introduced by the hydrodynamic model depends only on the gap width and triangle side length. Stronger second-harmonic transmittance correlates with stronger first harmonic transmittance, suggesting that ultranarrow gaps (below 10 nm) may not be the best candidate structures to observe SHG, despite exhibiting larger field enhancements due to the increased confinement.

\begin{figure}[h!]
 \centering
 \includegraphics[scale = 1.25]{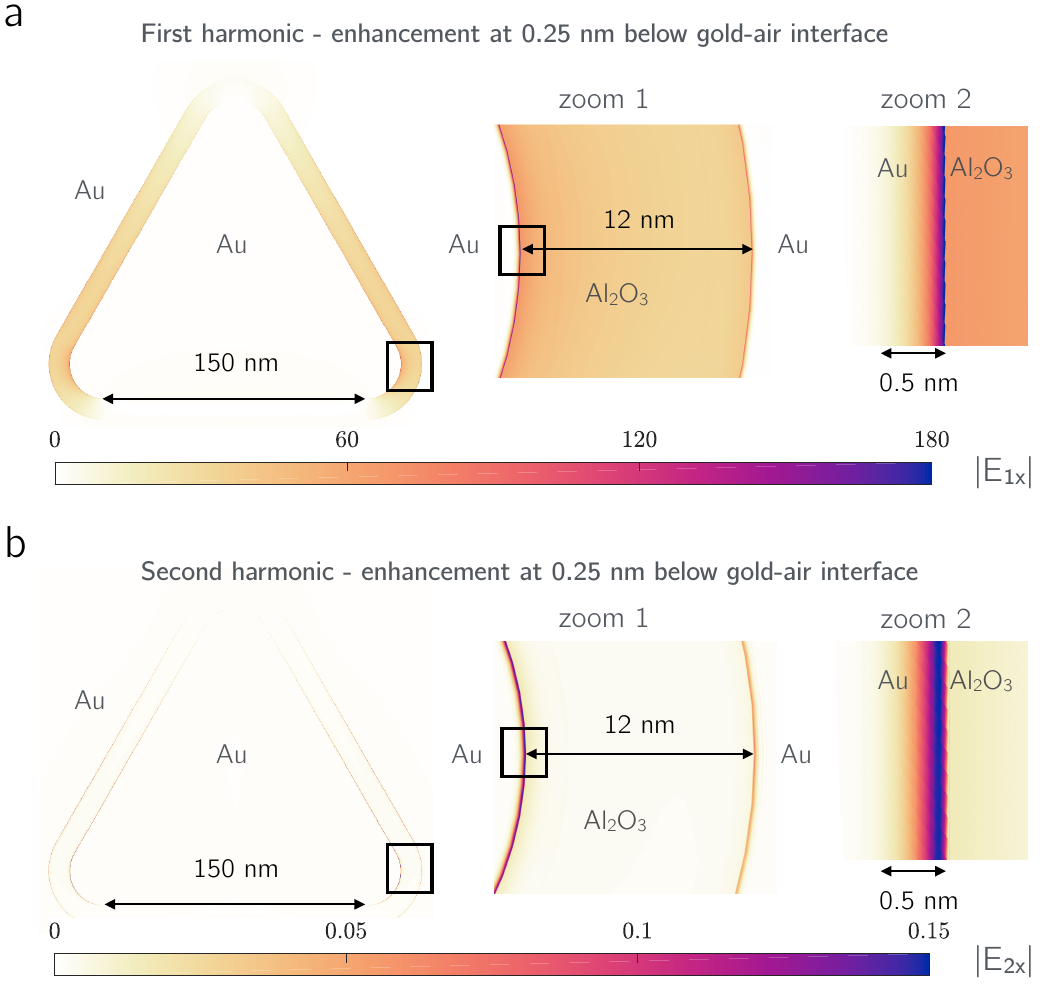}
 \caption{Cross-section image, with two zooms to show detail at the rounded vertex, of $x$-component of electric field for $D=150$ nm, $G=12$ nm and $T=128$ nm (double resonant structure) at resonant wavelength $\lambda = 2.009$ micron, computed 0.25 nm below the gold-air interface. Black squares indicate area of zoom. (a) First harmonic $\abs{\rm E_{1x}}$ (b) Second harmonic $\abs{\rm E_{2x}}$ .}\label{fig:modes}
\end{figure}

In order to highlight the importance of having a doubly resonant structure for enhanced SHG, we compute for $D=150$ nm and $G=12$ nm the transmittance profile for several film thicknesses, namely 110, 120, 128, 140 and 150 nm, shown in Figs. \ref{fig:shg_response}(c) and (d). For this gap and triangle side, thicknesses of approximately 128 nm are the ones that guarantee double resonances, as shown in Fig. \ref{fig:heatmap}(a) (top), thus we can expect the second harmonic transmittance $\pow_2$ to peak for this geometric configuration. Indeed, when comparing the peak transmittance of 120 and 128 nm thicknesses, we see that despite exhibiting lower first harmonic transmittance $\pow_1^{128} / \pow_1^{120} = 0.95$, the 128 nm thickness attains a second harmonic transmittance nearly four times larger than that of 120 nm $\pow_2^{128}/\pow_2^{120} = 3.97$. This boosting in second harmonic transmittance is a consequence of an optimal geometric configuration that excites resonances at precisely $\omega$ and $2\omega$. The field enhancement for this optimal geometry at the resonant wavelength is shown in Fig. \ref{fig:modes}, where the extreme confinement and boundary-layer structure of both the first and second harmonic at the curved metal-alumina interface can be appreciated. 


\section{Conclusions} \label{sec:conc}
The hybridizable discontinuous Galerkin method for Maxwell's equations augmented with the hydrodynamic model for metals is specially suited to simulate nonlinear plasmonics phenomena, owing to its high-order accuracy and its ability to handle the very large disparity in length scales and the extreme localization of electromagnetic fields.  For complex structures the required spatial discretization gives rise to a system that cannot be directly solved due to storage limitations. On the other hand we have not found a robust and effective iterative algorithm to solve the large indefinite HDG linear systems. In this article, we have presented a computational strategy to efficiently solve linear systems of equations that arise from the HDG method by performing a nested hybridization. In computational terms, we consider  discretizations that results form the extrusion of 2-D meshes and substitute one single large linear system solve for multiple smaller linear systems that stem from partitioning the original mesh into non-overlapping blocks of mesh elements following the extrusion direction. Furthermore, a judicious block partition enables us to reuse computations, thus making the nested HDG more efficient than classical HDG both from the storage and the computational perspective.


\section*{Acknowledgements}
F.~V.-C., N.-~C.~N and J.~P. acknowledge support from the AFOSR Grant No. FA9550-19-1-0240. S.-H.O. acknowledge support from the NSF Grant No. ECCS 1809240 and ECCS 1809723. F.~V.-C. acknowledges Vimworks for the design of Fig. \ref{fig:triangle} (a).

\bibliography{mainbib}
\bibliographystyle{elsarticle-harv}

\end{document}